\documentclass{aa}  
\usepackage{lscape}
\usepackage{graphicx}
\usepackage{natbib}
\bibpunct{(}{)}{;}{a}{}{,}

\usepackage{txfonts}
\begin{document}
   \title{Physical Properties of Southern Infrared Dark Clouds\thanks{
          Based on observations made with the ESO 37-channel bolometer array SIMBA at
          the SEST telescope on La Silla, under programme ID 71.C-0633. }}

   \author{T. Vasyunina\inst{1}\thanks{Member of the International Max Planck Research School (IMPRS) Heidelberg}, H. Linz\inst{1}, Th. Henning\inst{1}, 
           B.~Stecklum\inst{2}, S.~Klose\inst{2}, L.--\AA. Nyman\inst{3}}

   \offprints{T.~Vasyunina}

   \institute{Max Planck Institute for Astronomy (MPIA),
              K\"onigstuhl 17, D-69117 Heidelberg, Germany\\
              \email{[vasyunina,linz,henning]@mpia.de} 
	  \and   
              Th\"uringer Landessternwarte Tautenburg,
	      Sternwarte 5, D - 07778 Tautenburg, Germany\\\
	      \email{[stecklum, klose]@tls-tautenburg.de}
	  \and  
	      ESO, Santiago 19, 19001 Casilla, Chile \\
	      \email{lnyman@eso.org}}

   \date{Received ; accepted }

% \abstract{}{}{}{}{} 
% 5 {} token are mandatory
 
  \abstract
   {What are the mechanisms by which massive stars form? What are the initial conditions
   for these processes? 
   It is commonly assumed that cold and dense Infrared Dark Clouds (IRDCs) likely represent the birth sites 
   massive stars. Therefore, this class of objects
   gets increasing attention, and their analysis offers the opportunity to tackle the above mentioned questions.}
   {To enlarge the sample of well-characterised IRDCs in the 
   southern hemisphere, where ALMA will play a major role in the near future, 
   we have set up a program to study the gas and dust of southern 
   infrared dark clouds.
   The present paper aims at characterizing the continuuum properties of this
   sample of IRDCs.}
   {We cross-correlated 1.2 mm continuum data from SIMBA@SEST with Spitzer/GLIMPSE images to 
   establish the connection between emission sources at millimeter wavelengths and the IRDCs 
   we see at 8 $\mu$m in absorption against the bright PAH background. Analysing the dust emission and extinction
   leads to a determination of masses and column densities, which are important quantities in 
   characterizing the initial conditions of massive star formation. We also evaluated 
   the limitations of the emission and extinction methods.  }
   {The morphology of the 1.2 mm continuum emission is in all cases in close agreement with the mid-infrared  
   extinction. The total masses of the IRDCs were found to range from 150 to 1150 $\rm M_\odot$ 
   (emission data) and from 300 to 1750 $\rm M_\odot$ (extinction data). We derived peak column 
   densities between 0.9 and 4.6 $\times  10^{22}$ cm$^{-2}$ (emission data) and 
   2.1 and 5.4 $\times  10^{22}$ cm$^{-2}$ (extinction data). We demonstrate that the extinction method fails
   for very high extinction values (and column densities) beyond A$_{\rm V}$ values of roughly 75 mag according to
   the Weingartner \& Draine (2001) extinction relation $R_{\rm V} = 5.5$ model B (around 200 mag when following the common
   Mathis (1990) extinction calibration). By taking the spatial resolution effects into account and restoring
   the column densities derived from the dust emission back to a linear resolution of 0.01 pc, peak column densities
   of 3\,--\,19 $\times  10^{23}$ cm$^{-2}$ are obtained, much higher than typical values for low-mass cores.}
   {The derived column densities, taking into account the spatial resolution effects, are beyond the column density
    threshold of 3.0 $\times  10^{23}$ cm$^{-2}$ required by theoretical considerations for massive star formation. 
    We conclude that the values for column densities derived for  
    the selected IRDC sample  make these objects excellent candidates for objects in the
    earliest stages of massive star formation.}


   \authorrunning{T. Vasyunina et al.}
   \maketitle
%
%________________________________________________________________

\section{Introduction}

One of the key questions in stellar astrophysics is to understand the formation and earliest
evolution of high-mass stars. These objects play a major role in shaping the 
interstellar medium due to their strong UV radiation fields and stellar winds, 
and they enrich their environment with heavy elements when
exploding as supernovae. Despite their importance, the 
mechanism by which such massive stars form and, especially, the initial conditions 
of their birthplaces are poorly understood \citep{2007prpl.conf..165B,2007ARA&A..45..481Z}.

One pathway to tackle especially the latter problem is to analyse so-called 
Infrared Dark  Clouds (IRDCs). They were first identified with the 
\emph{Infrared Space Observatory \/} \citep[\emph{ISO\/};][]{1996A&A...315L.165P} and
\emph{Midcourse Space Experiment\/} \citep[\emph{MSX\/};][]{1998ApJ...494L.199E} 
as dark extended features against the bright Galactic PAH background at
mid-IR (MIR) wavelengths. The first studies 
\citep{1998ApJ...494L.199E,1998ApJ...508..721C,2000ApJ...543L.157C} showed that 
IRDCs are dense ($>$10$^5$ cm$^{-3}$), cold ($<$25 K) and can attain high column
densities ($\ga$ 10$^{23}$ cm$^{-2}$). All these properties 
make IRDCs excellent candidates for hosting very early stages of massive 
star formation.

During the last years, additional studies of Infrared Dark Clouds at
millimeter and submillimeter wavelengths were performed.
\citet{2006ApJ...639..227S} 
presented a catalog of almost 11\,000 IRDCs in the first and fourth quadrants 
of the Galactic plane. Using $^{13}\rm CO$ J=1--0 molecular line emission, 
the kinematic distances to 313 clouds from this catalog were established
\citep{2006ApJ...653.1325S}. This allowed to estimate sizes, masses and the
Galactic distribution for this large sample. The study showed that IRDCs have 
sizes of $\sim$ 5 pc and LTE masses of $\sim$ 5 $ \times $ 10$^3$ $\rm M_\odot$
comparable to cluster-forming molecular
clumps. The galactic distribution of IRDCs follows the general distribution of molecular gas. A
concentration of the clouds is associated with the so-called 5 kpc molecular ring, the
Galaxy's most massive and active star-forming complex.
Ammonia observations of some well-known IRDCs with the Effelsberg 100 m telescope 
\citep{2006A&A...450..569P} allowed to estimate additional chemical and 
physical properties, such as average gas temperature (between 10 and 20 K), 
velocity fields (significant velocity gradient between the cores, 
linewidths of 0.9\,--\,1.5 km/s) and the chemical state. According to this study, $\rm NH_3$ in IRDCs 
is overabundant by a factor of 5\,--\,10 relative to Taurus or Perseus local dark clouds, 
while $\rm H_2CO$ is underabundant by a factor of $\sim$ 50. Hence, the chemistry 
governing IRDCs might be complex and could be different from other parts of the molecular ISM.

Although significant progress has been made observationally, 
the number of IRDCs with well characterised properties is still 
small to date, especially regarding the southern hemisphere.
To enlarge the sample of well-investigated IRDCs, we selected 12 clouds in 
the southern hemisphere and started a program to measure 
the gas and dust properties of these objects. 

In Sect. 2 we describe our source selection and 1.2 mm continuum observations
with the SIMBA/SEST telescope. In Sect. 3, we discuss the data reduction and
the details concerning the calculation of dust masses and column densities. Also we present here 
the comparison of the MIR and millimeter techniques.
In Sect. 4, we compare our results with previous results for high- 
and low-mass star-forming region and  with the theoretical models.

\section{Observations}

  The IRDCs for our study were selected in the "pre--Spitzer" era, by visual examination 
  of the MIR images delivered by the MSX
  satellite.  The MSX A band (6.8 -- 10.8
  $\mu$m) was the most sensitive one among the MSX bands and presented
  the highest level of diffuse background emission (due to
  PAH emission at 7.7 and 8.7 $\mu$m), which also leads to the highest
  contrast between bright background and dark IRDCs. We selected
  a sample of southern IRDCs from the A band images looking for high contrast
  and sizes sufficient to fill the main beam of the SEST telescope at 1.2 mm. \\
  The 1.2 mm continuum observations were carried out with the
37-channel bolometer array SIMBA \citep{2001Msngr.106...40N} at
the SEST on La Silla, Chile between July 16\,--\,18, 2003. SIMBA is a hexagonal array in
which the HPBW of a single element is about 24$''$ and the
separation between elements on the sky is 44$''$. The observations
were made using a fast mapping technique without a
wobbling secondary \citep{2002A&A...383.1088W}. 

Maps of Uranus were taken to check the flux calibration of the resulting data. 
To correct for the atmospheric opacity, skydips were performed every
2\,--\,3 hours. Despite the occurrence of some thin clouds, the observing
conditions were good which is reflected in zenith opacity
values of 0.16\,--\,0.18. The pointing was checked roughly every
two hours and proved to be better than 6$''$. The combination of typically 
three maps  with sizes of 560$'' \times 900''$ resulted in a residual
noise of about 22\,--\,28 mJy/beam (rms) in the center of the mapped region.

%__________________________________________________________________

\section{Data reduction and analysis}

In this paper we are using both 8\,$\mu$m IRAC data from the Spitzer  
Galactic Legacy Infrared Mid-Plane Survey Extraordinaire \citep[GLIMPSE,][]{2003PASP..115..953B} 
and our 1.2 mm data from the SIMBA bolometer at the SEST telescope
to investigate the physical properties of the extinction and emission material. 

In both cases, for 
deriving the masses of the IRDCs, we need a handle on the distances to these clouds.
For determining the (kinematic) distances to our IRDCs, we use the $v_{\rm LSR}$ velocities derived from 
recent molecular line observations\footnote{The HCO$^+$(1-0) line velocities have been employed for
this purpose.} with the Australian MOPRA telescope, which we will present in a forthcoming paper. The
velocities have been transfered to kinematic distances by adopting the recently improved parameters for the 
Galactic rotation curve \citep{2008ApJ...679.1288L} for the fourth and first Galactic quadrant. Always the near
kinematic distance has been assumed. The corresponding distances are reported in Table \ref{table:1} and
have been used for the mass estimations.
Note that such rotation curves give just average properties. The actual distribution of material 
might be more structured, especially in the fourth quadrant, which is indicated in the H{\sc I} absorption 
measurements shown in \citet{2008ApJ...679.1288L}. Furthermore, for objects in the Galactic 
longitude interval [$305^\circ,310.5^\circ$], several velocity systems can occur due to the projection of at 
least two Galactic arms. The measured velocities, however, place all our IRDCs in that longitude range within
the Centaurus arm (3.5\,--\,5.5 kpc), in agreement with \citet{2001PASJ...53.1037S}.

\subsection{Millimeter data}\label{Sect:SIMBA-Data}

The 1.2 mm data for the IRDC regions from SIMBA at the SEST telescope were 
reduced using the MOPSI package (developed by R.~Zylka, IRAM). All maps were 
reduced by applying the atmospheric opacity corrections, fitting and subtracting 
a baseline, and removing the correlated sky noise. Thereby, we followed a three-stage
approach as suggested in the SIMBA manual. After a first iteration using all data for the
sky noise removal, the map regions showing source emission are neglected for sky noise removal 
in the second iteration. From this second interim map a source model is derived which is being 
included in the third iteration.  The resulting maps were flux--calibrated using 
the conversion factor obtained from  observations of Uranus. For our July 2003 observations, this
factor was around 60 mJy/beam per count.

For estimating cloud masses and column densities we used the following 
expressions: %\citep{2002ApJ...566..945B,2005ApJ...633..535B}:
\begin{equation}\label{Equ:dustmass}
M_{\rm tot} =   \frac {F_{\rm int} \ D^2 \ R}{B_v(T) \ \varkappa_v} ,
\end{equation}

\begin{equation}\label{Equ:columndens}
N_{\rm H_2} =   \frac {F_{\rm peak} \ R}{\Omega \ B_v(T) \ \varkappa_v \ {\rm m}_{\rm H_2}} .
\end{equation}

$F_{\rm peak}$ is the measured source peak flux density, while $F_{\rm int}$ denotes the integrated flux density
of the whole source. $\Omega$ is the beam solid angle in steradians, m$_{\rm H_2}$ is the mass of one hydrogen
molecule, $D$ is the distance to the IRDC, $R$ is the gas-to-dust ratio, $\varkappa_v$ is the dust opacity per 
gram of dust, and $B_v(T)$ is the Planck function at the dust temperature $T$. We adopt a gas-to-dust mass 
ratio of 100, and $\varkappa_v$ equal to 1.0 cm$^2$ g$^{-1}$, a value appropriate for cold dense cores 
\citep{1994A&A...291..943O}. 

At the present stage, where measured temperatures are not available for our sources, 
we assume a temperature of 20 K, which is a 
reasonable choice considering recent investigations toward other IRDCs 
\citep{1998ApJ...508..721C,2006A&A...447..929P}. \\
The derived mass depends on the (assumed) temperature,  
on the distance to the cloud, and on the grain model. 
Masses will be underestimated if the temperature is lower than the assumed 
value of 20 K. For example, for 15 K the masses will be higher by 
a factor of 1.5. In case of a higher temperature in the cloud, e.g., for 30 K, our
results have to be divided by a factor of 1.7.
We further note the
quadratic dependence of the derived masses on the distance of the clouds.
Hence, the masses will be by a factor of 1.2 - 1.8 higher if the distance is 500 pc more,
and lower by the same factor if it is 500 pc less than indicated by the average Galactic 
rotation curve (see above).
The derived masses are inverse proportional to the assumed value of the opacity $\varkappa_v$,
which has an uncertainty of at least a factor 2.
The column density has no direct dependence on the distance to the cloud, 
but the temperature dependence is the same as for the masses.

   \begin{figure*}
   \centering
   \includegraphics[width=8cm]{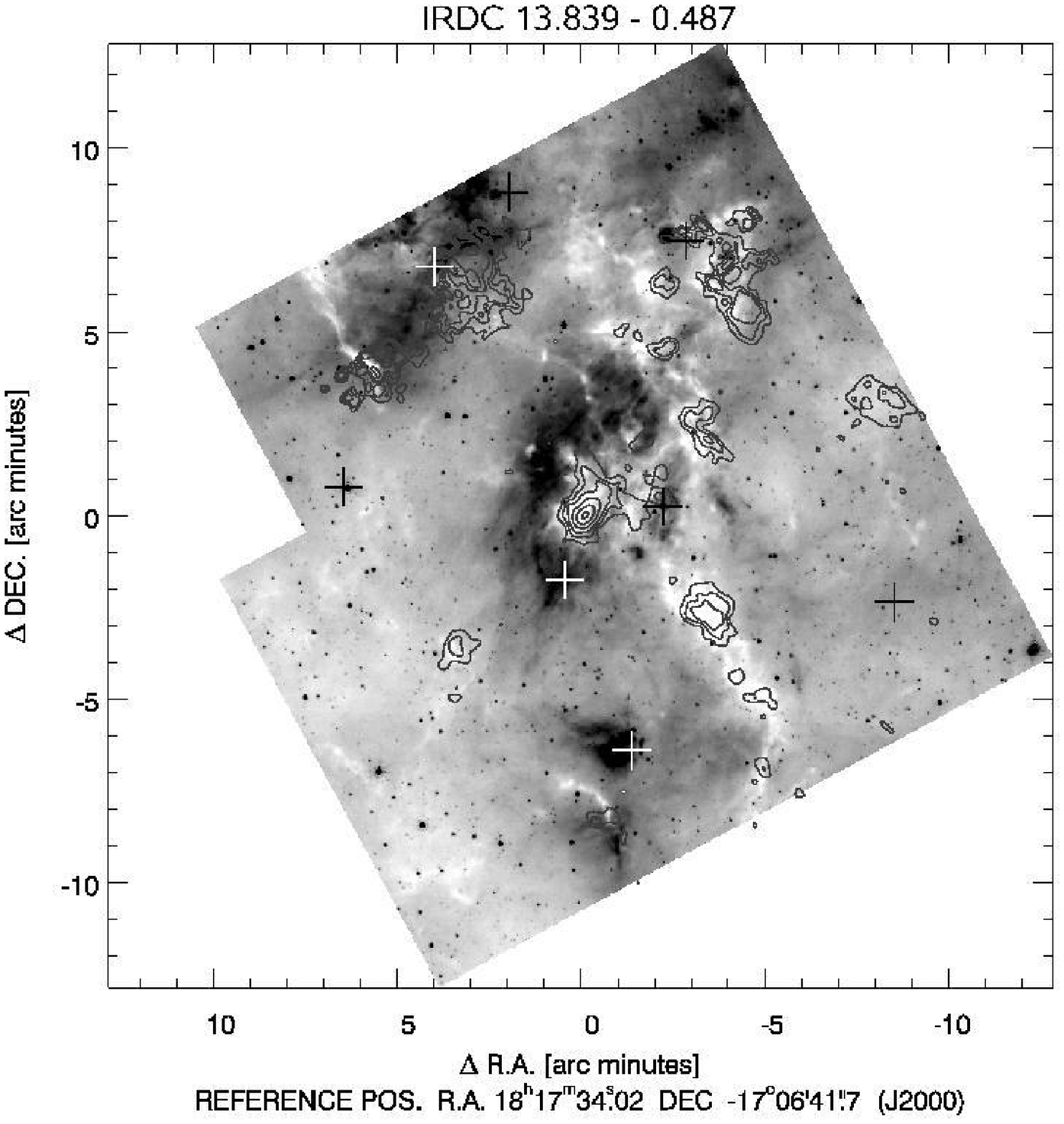}
   \includegraphics[width=7.9cm]{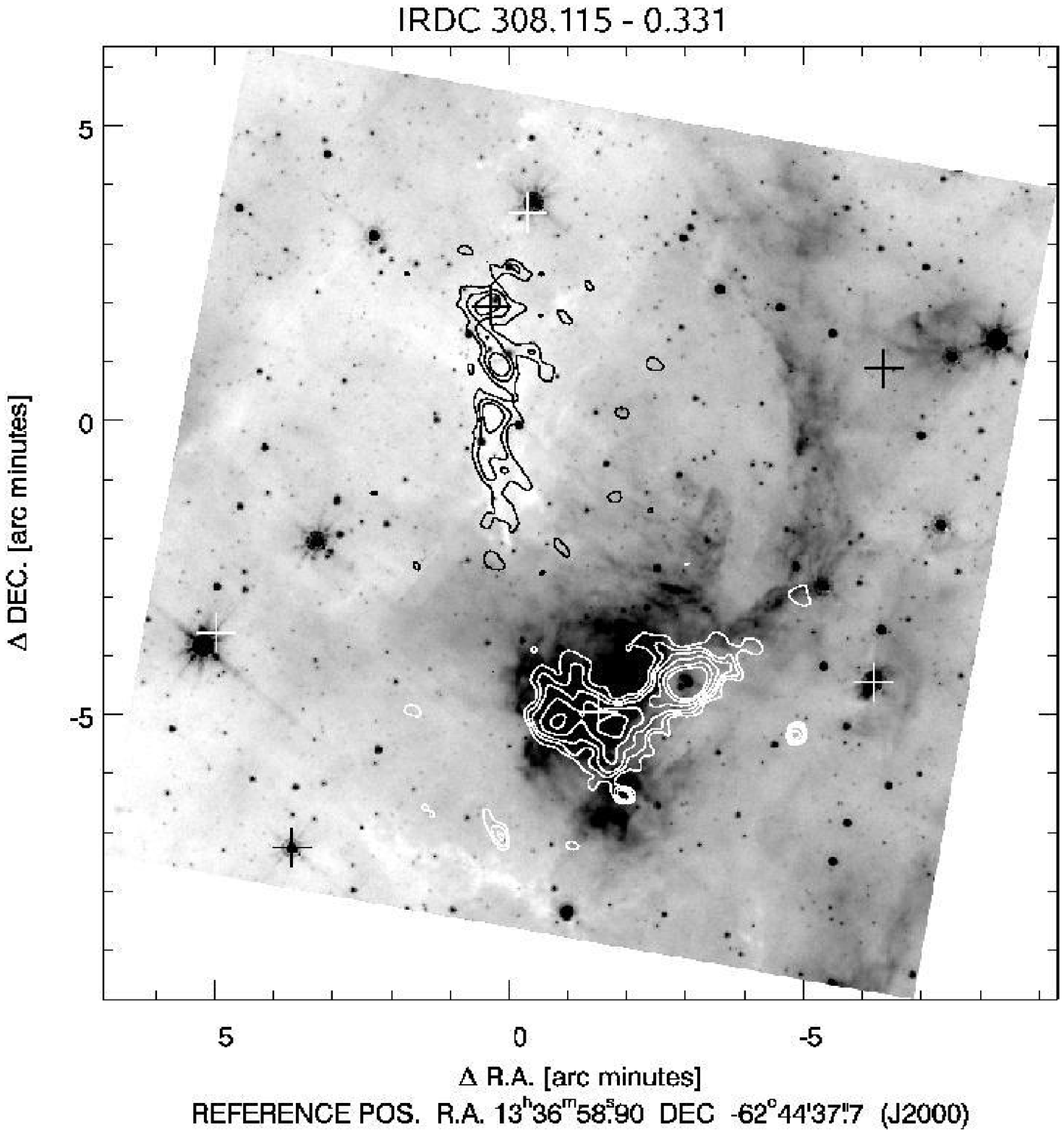}
   \includegraphics[width=8cm]{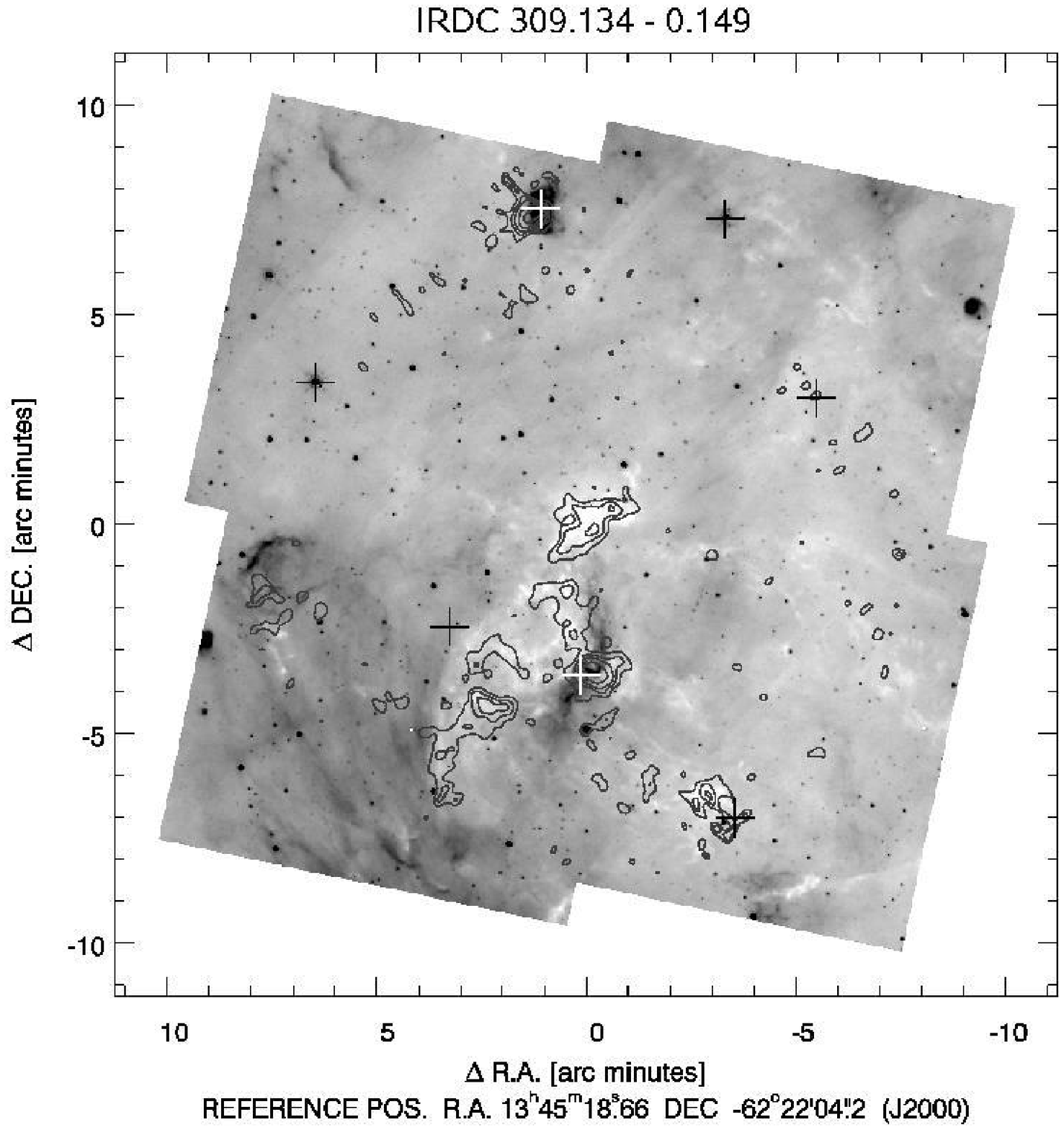}
   \includegraphics[width=8cm]{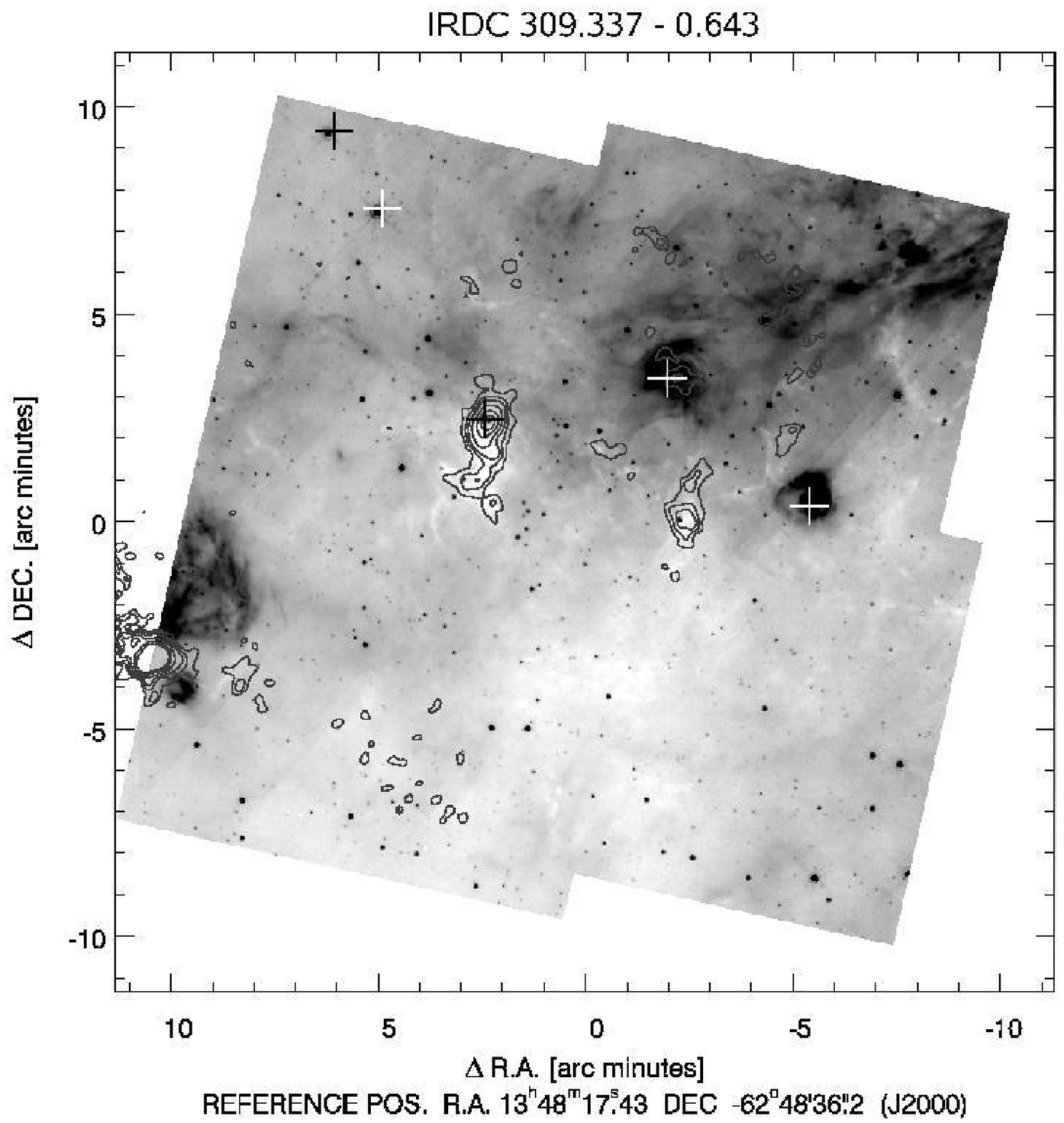}
   \includegraphics[width=8cm]{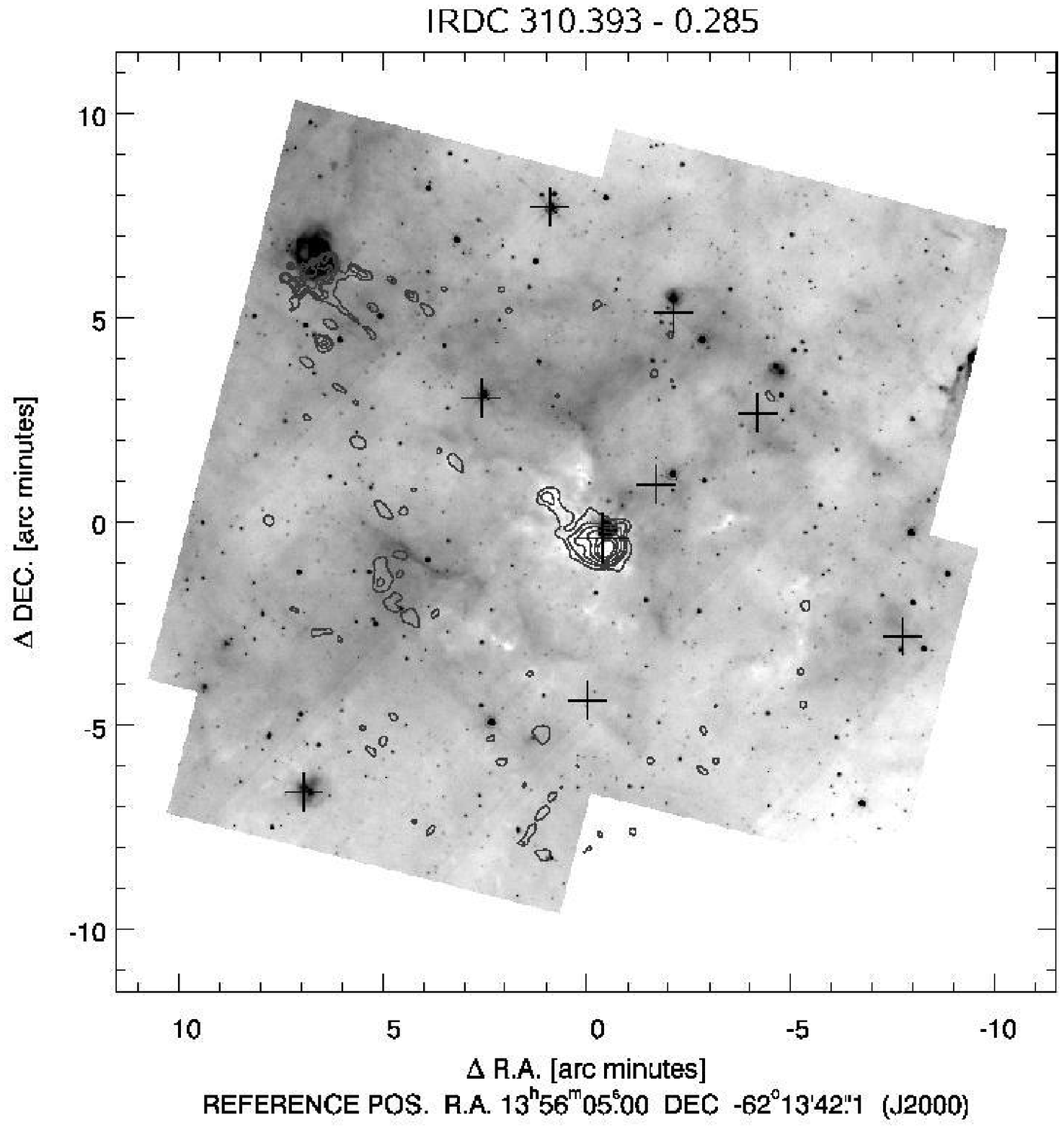}
   \includegraphics[width=7.9cm]{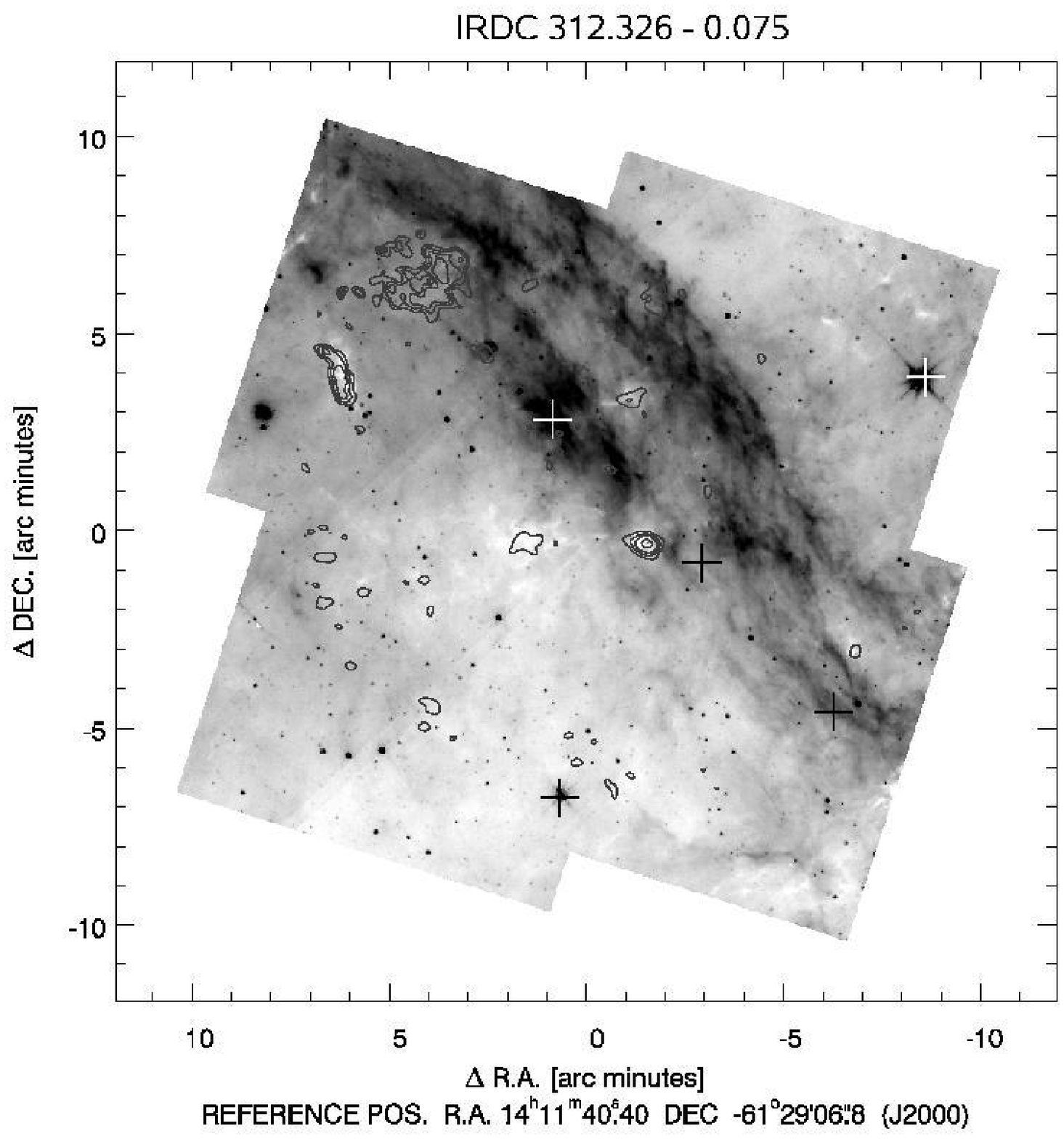}
   \caption{Inverse grey-scale 8 micron maps overlaid with 1.2 mm continuum emission.
   The intensity of a grey-scale image corresponds to the square root of the inverse intensity in mJy.
   The contours are 60, 108, 156, 240, 360, 480 mJy beam$^{-1}$ in all cases. Crosses
   denote the position of IRAS point sources.
              }
         \label{Fig1}
   \end{figure*}

   \begin{figure*}
   \centering
   \includegraphics[width=8cm]{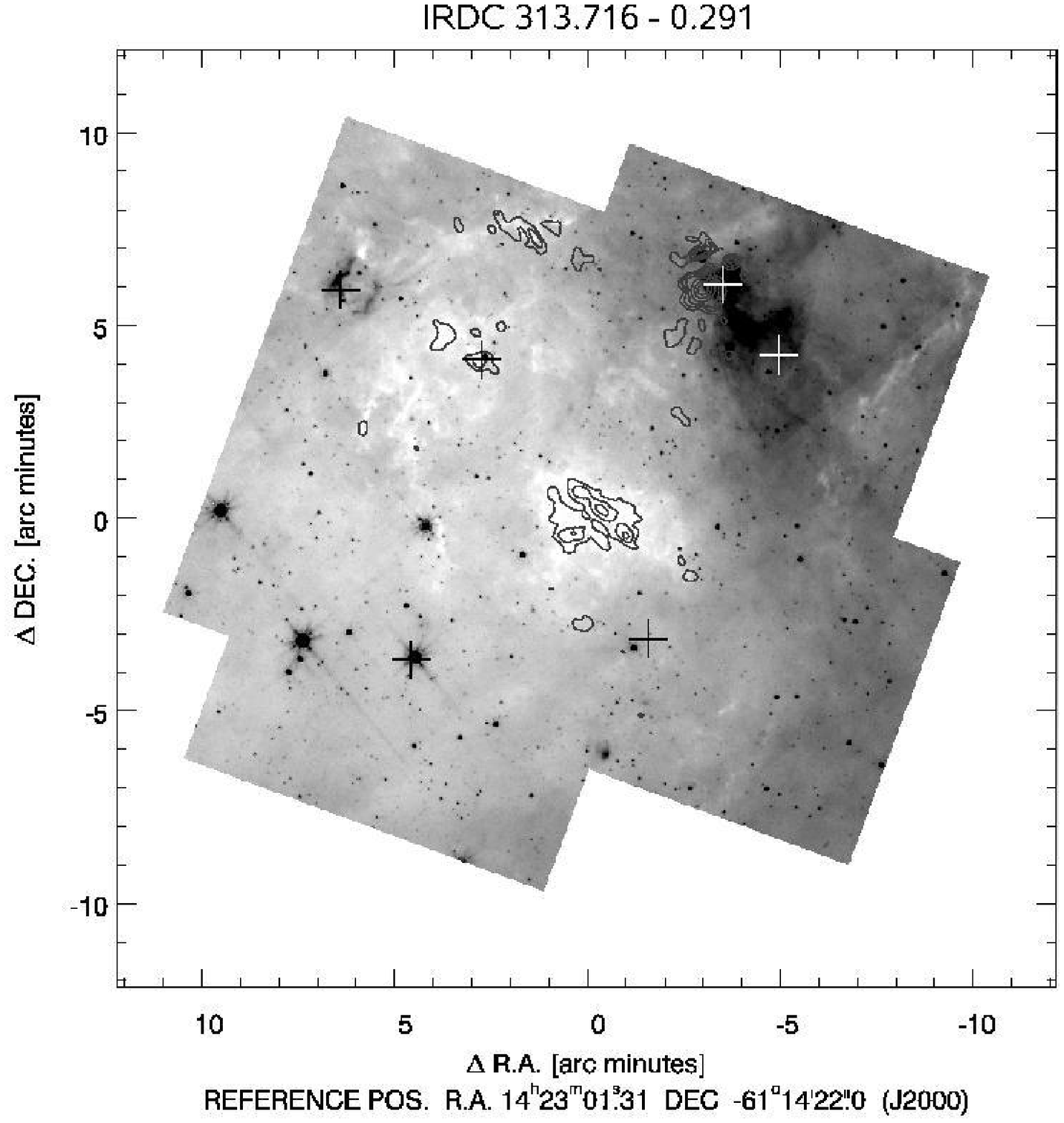}
   \includegraphics[width=8cm]{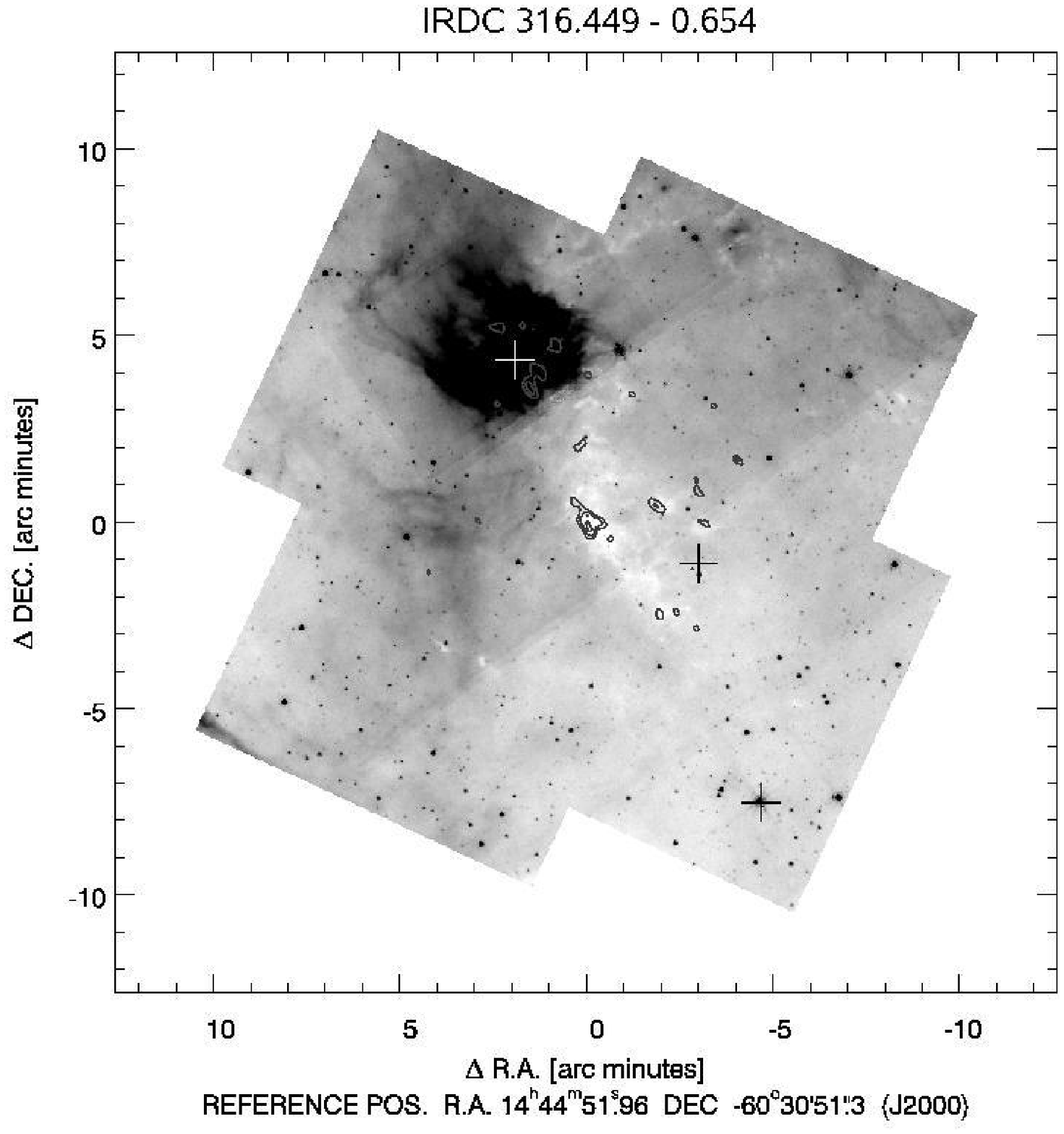}
   \includegraphics[width=8cm]{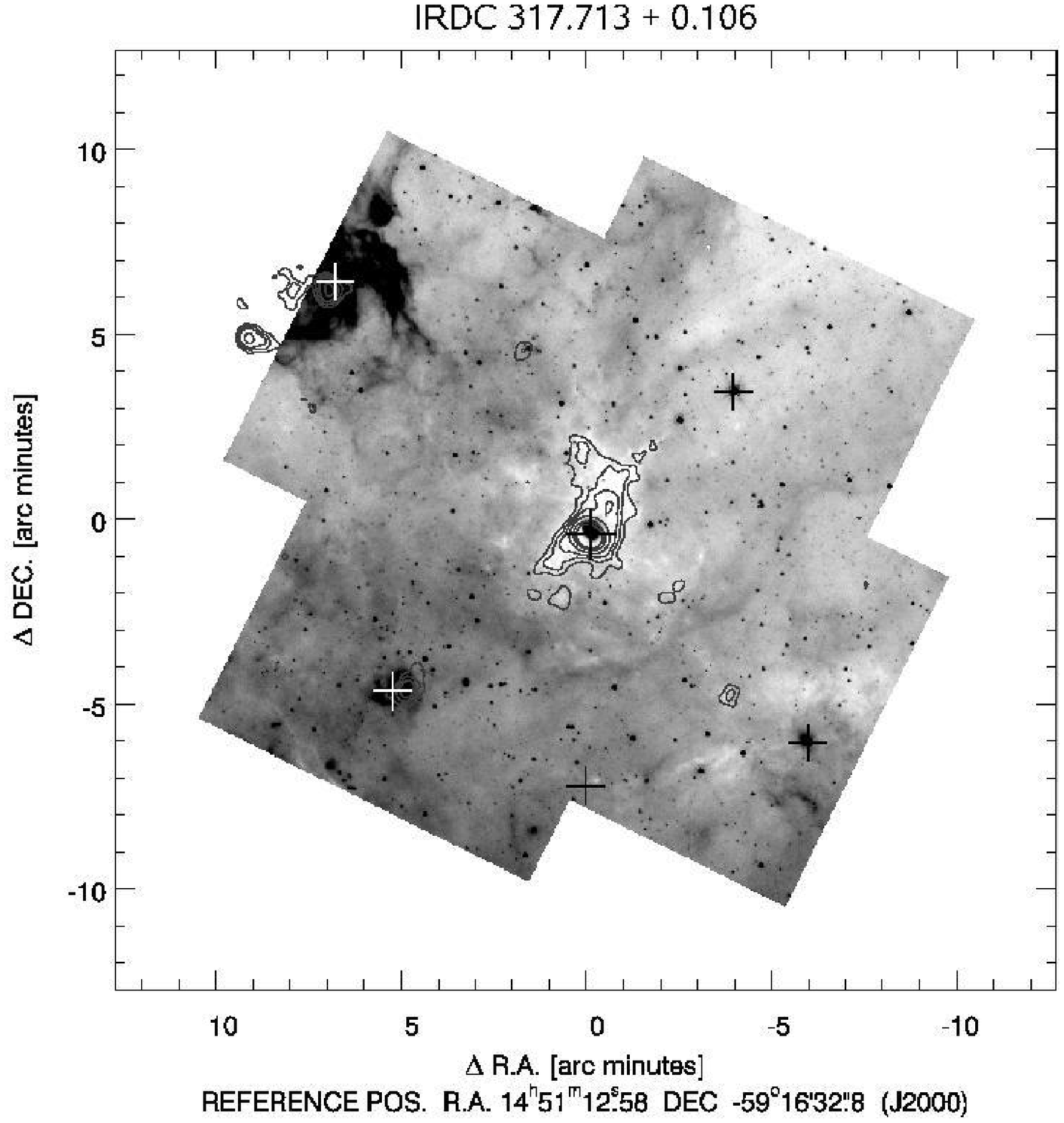}
   \includegraphics[width=8cm]{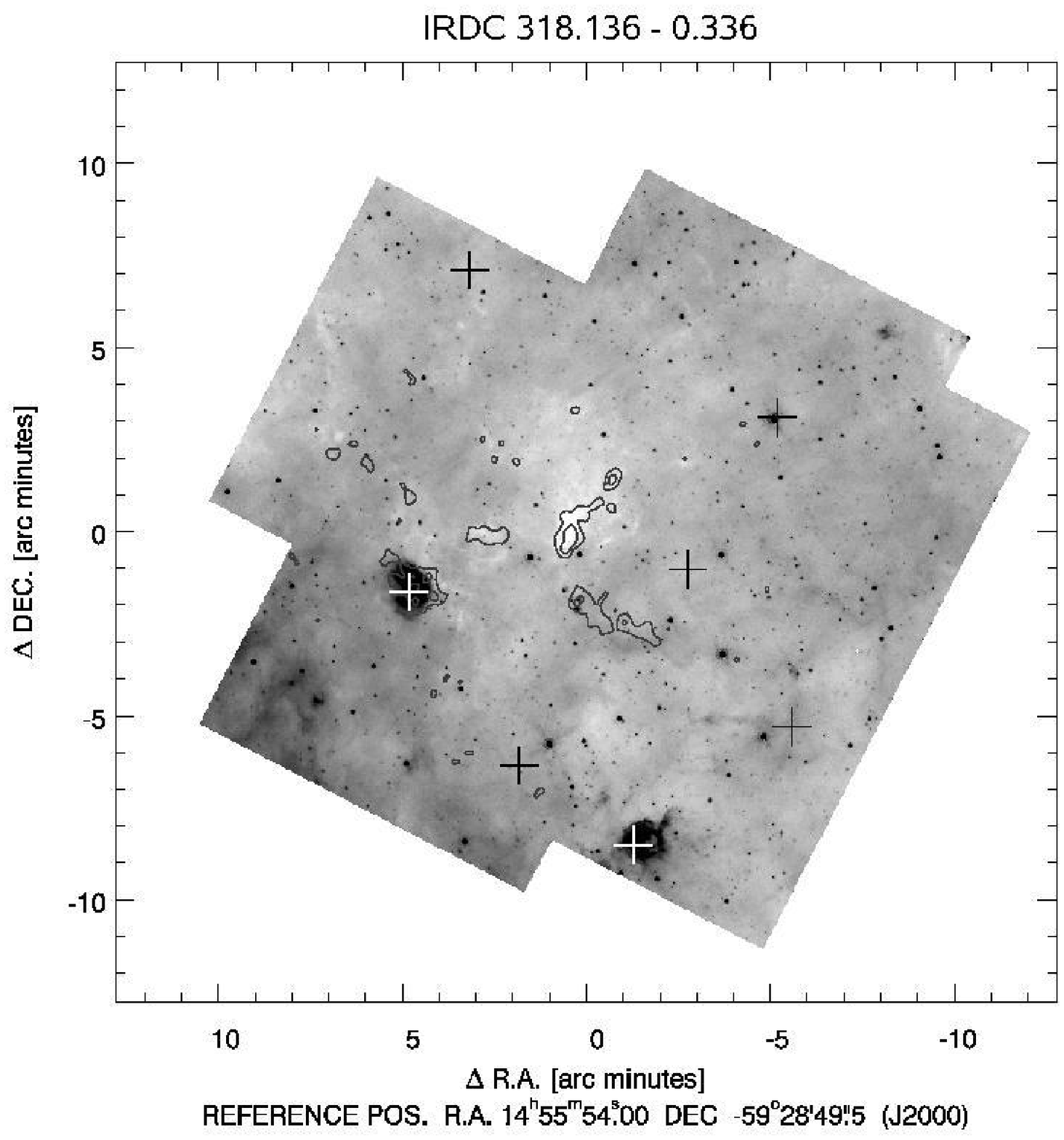}
   \includegraphics[width=8cm]{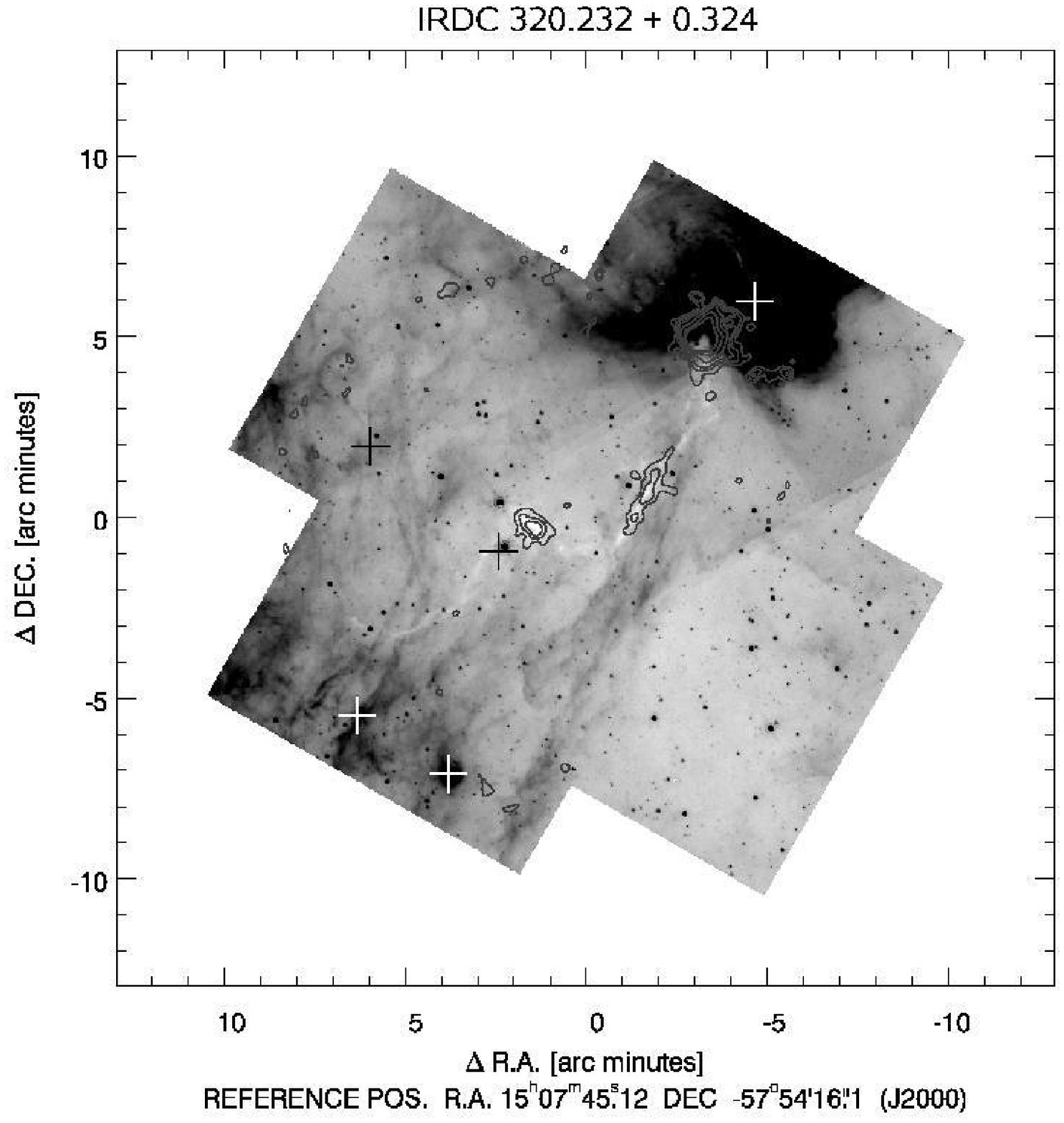}
   \includegraphics[width=8cm]{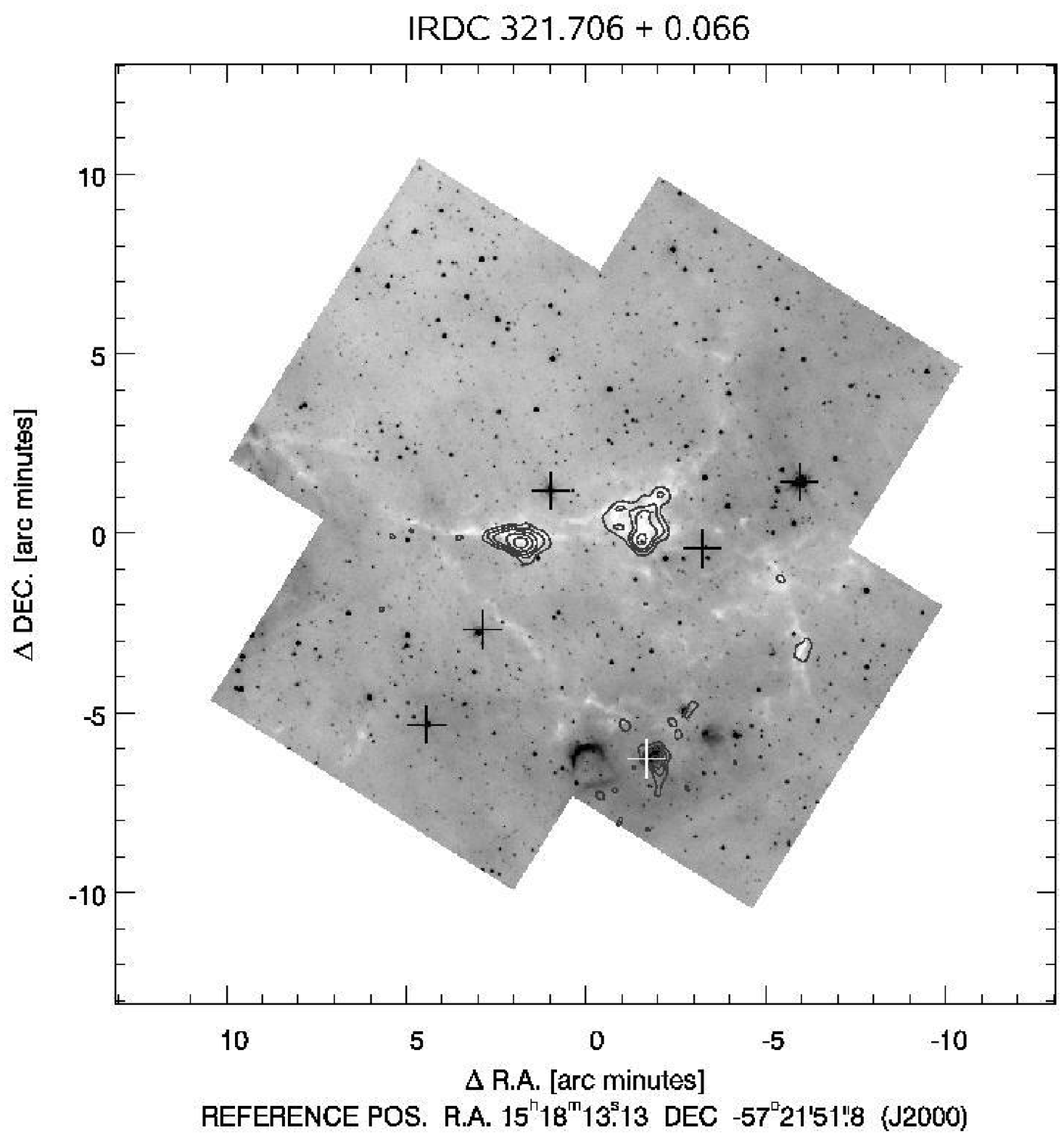}
   \caption{Inverse grey-scale 8 micron maps overlaid with 1.2. mm continuum emission.
   The intensity of a grey-scale image corresponds to the square root of the inverse intensity in mJy.
   The contours are 60, 108, 156, 240, 360, 480 mJy beam$^{-1}$ in all cases except
   for IRDC 316.45-0.65 where it is 84, 120, 156, 240, 360, 480 mJy beam$^{-1}$. Crosses
   denote the position of IRAS point sources.
              }
         \label{Fig2}
   \end{figure*}

   \begin{figure*}
   \centering
   \includegraphics[width=6.0cm]{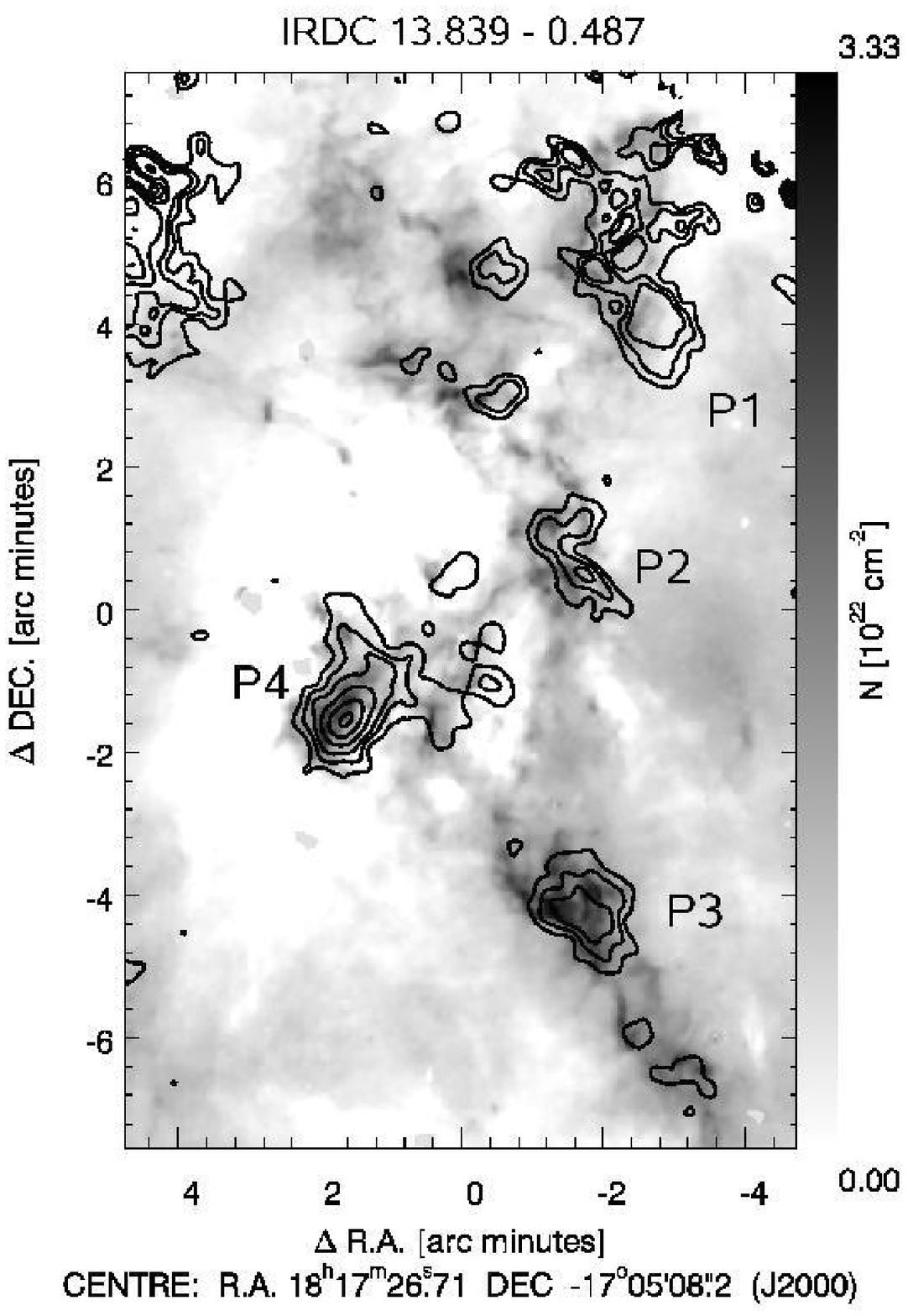}
   \includegraphics[width=6.0cm]{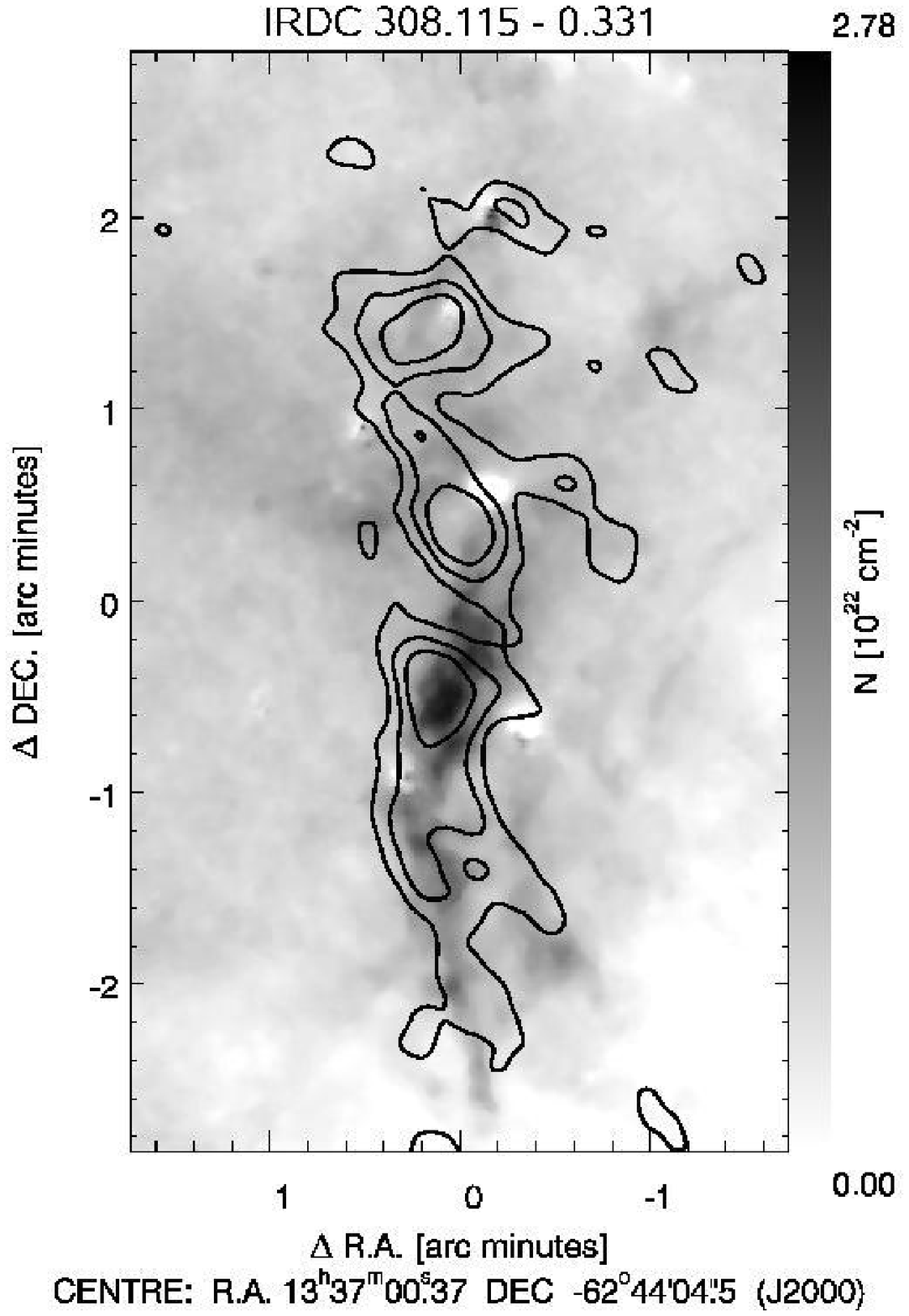}
   \includegraphics[width=7.0cm]{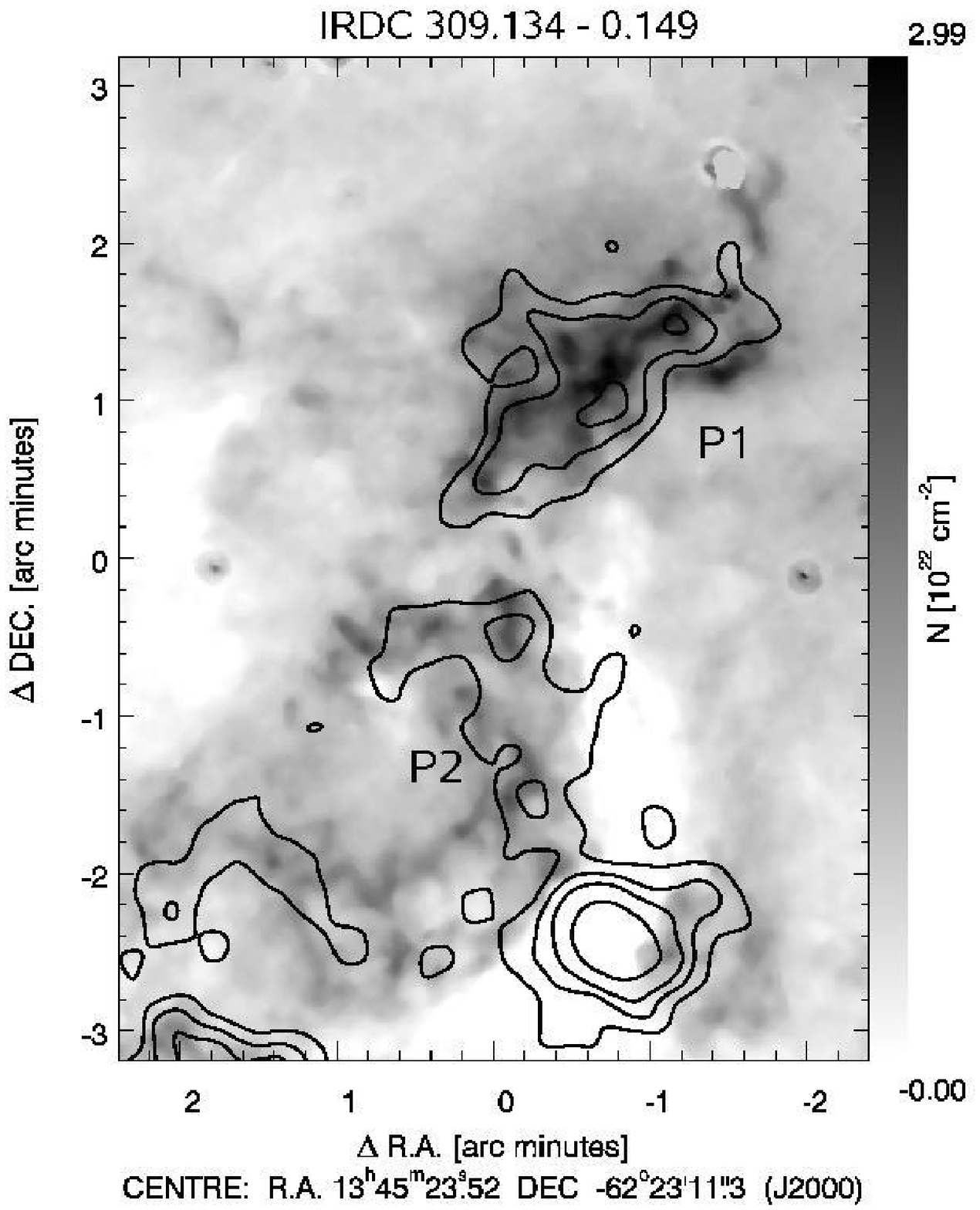}
   \includegraphics[width=8.5cm]{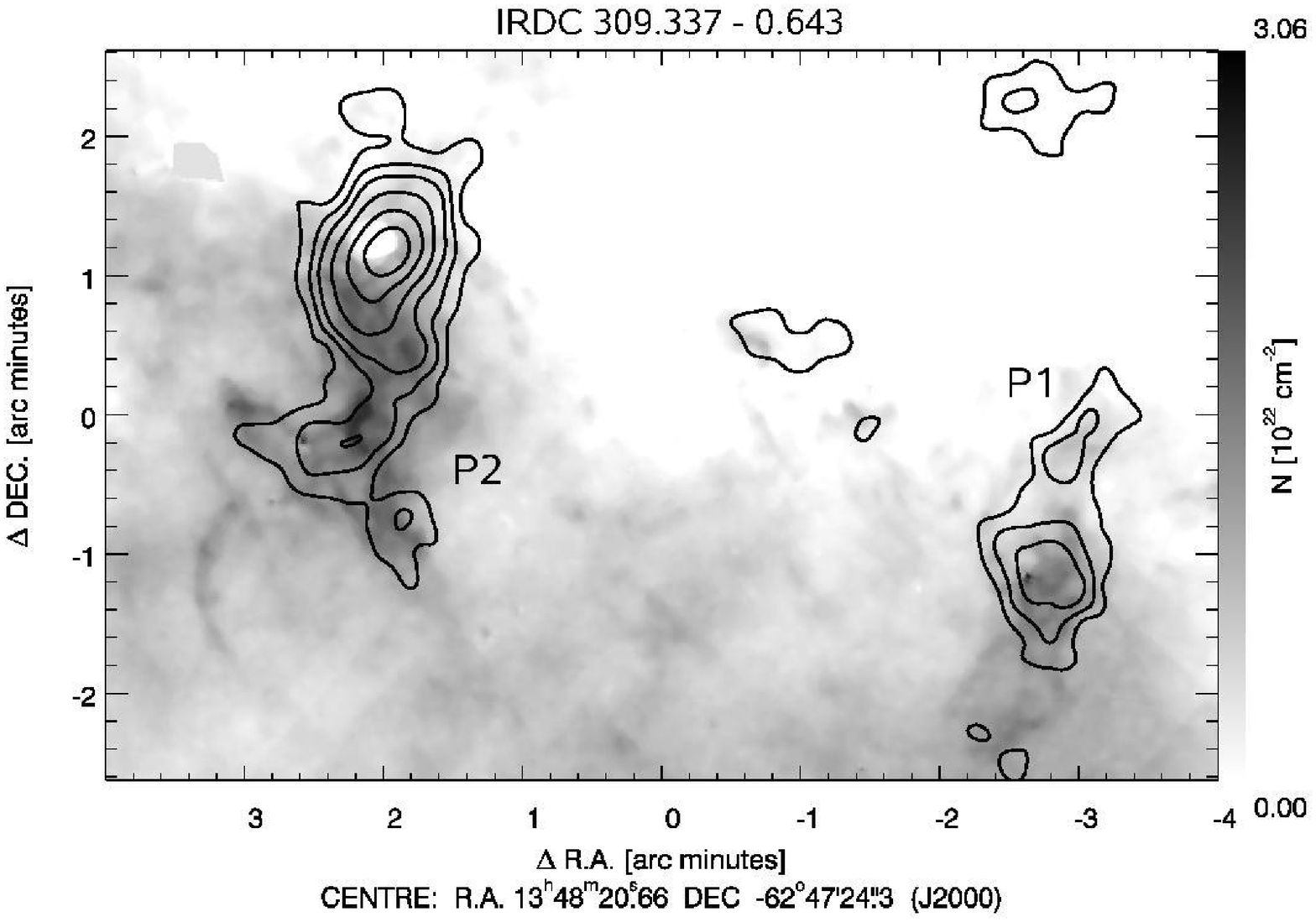}
   \includegraphics[width=7.5cm]{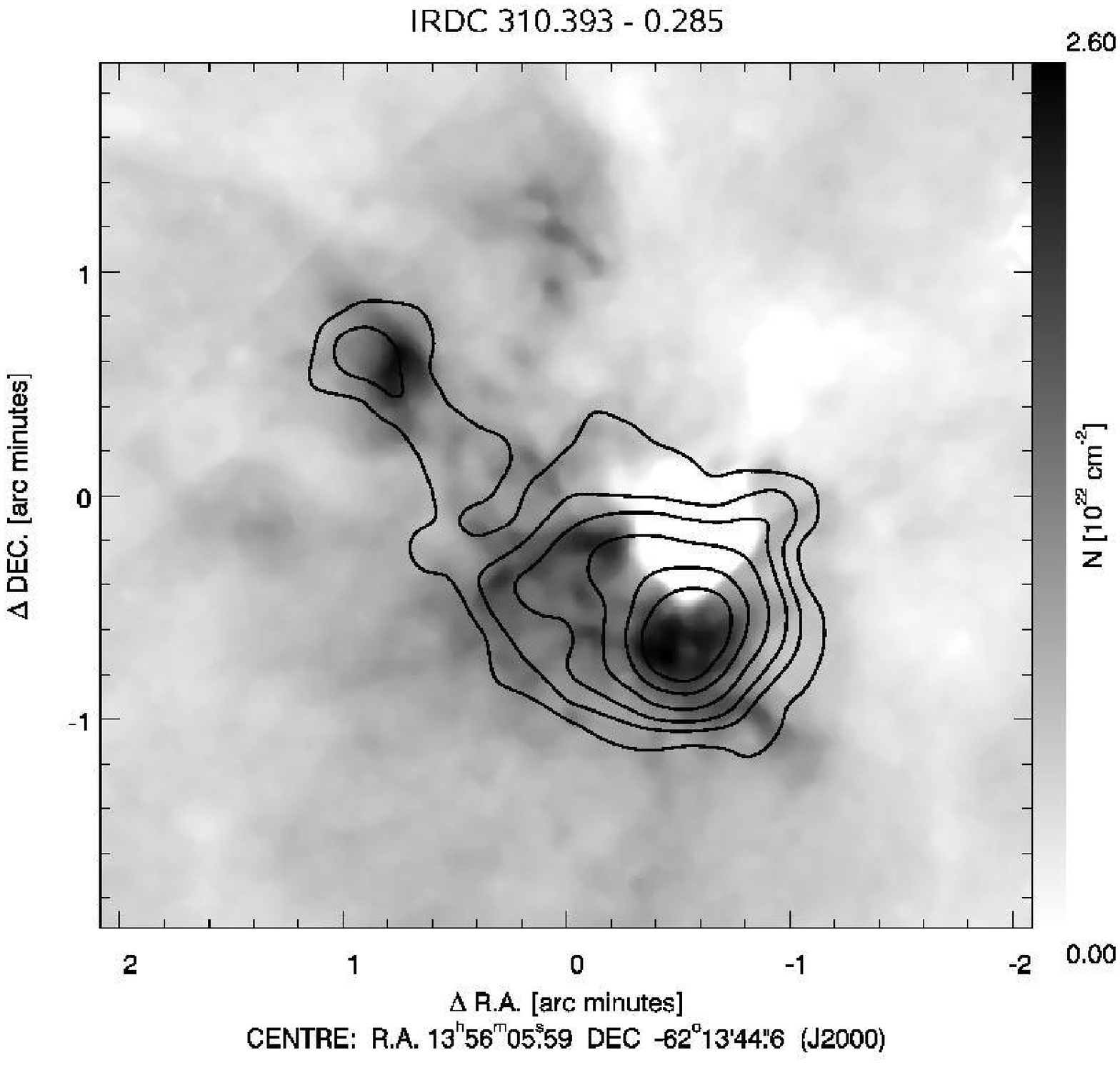}
   \includegraphics[width=8.0cm]{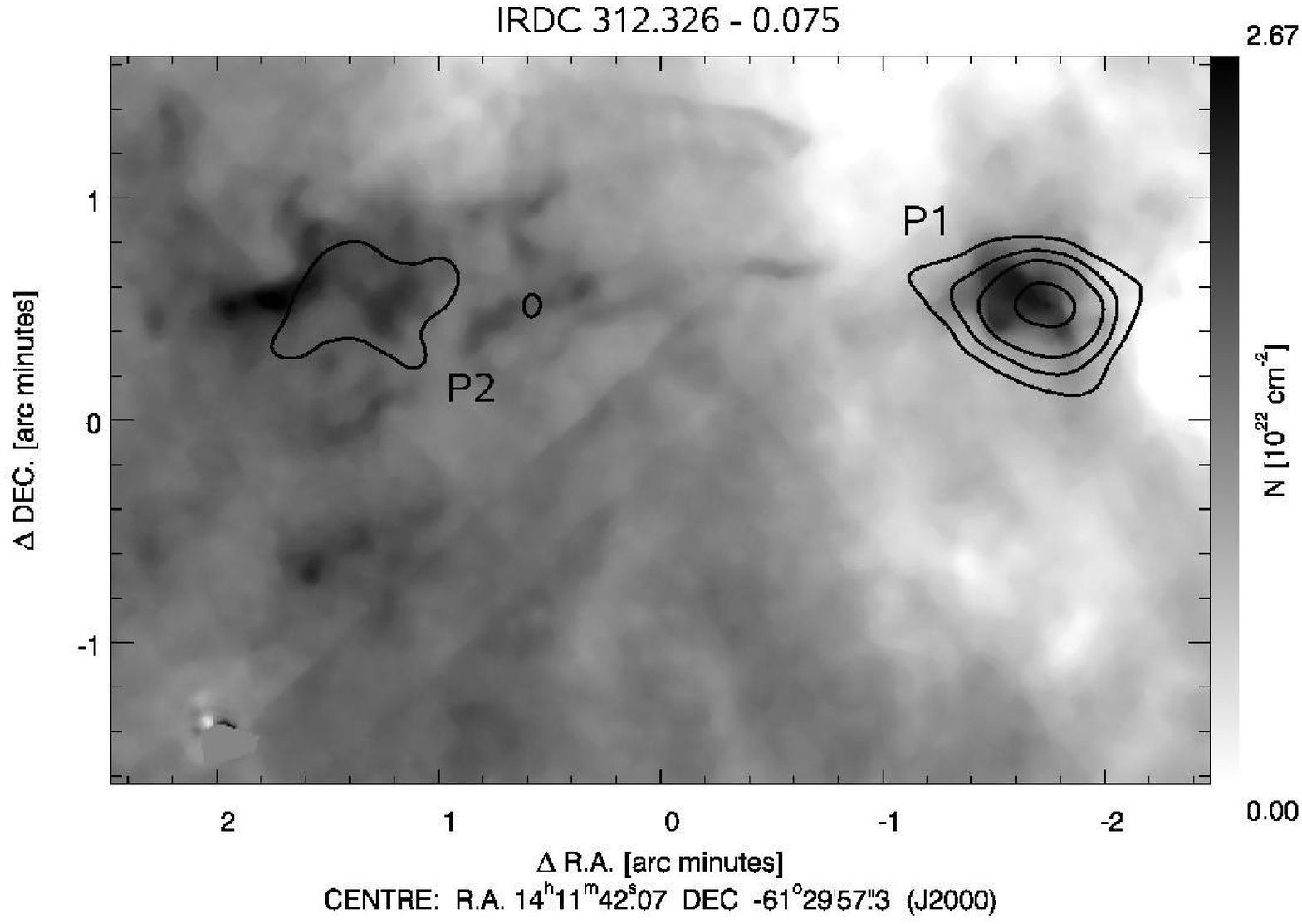}
   
   \caption{Column density maps derived from 8$\ \mu$m extinction overlaid with 1.2 mm 
   continuum emission as contours. The scaling is indicated in the bar to the right of each image.
   The contours are 60, 108, 156, 240, 360, 480 mJy beam$^{-1}$ in all cases.
              }
         \label{Fig3}
   \end{figure*} 
      
   \begin{figure*}
   \centering
   \includegraphics[width=8.3cm]{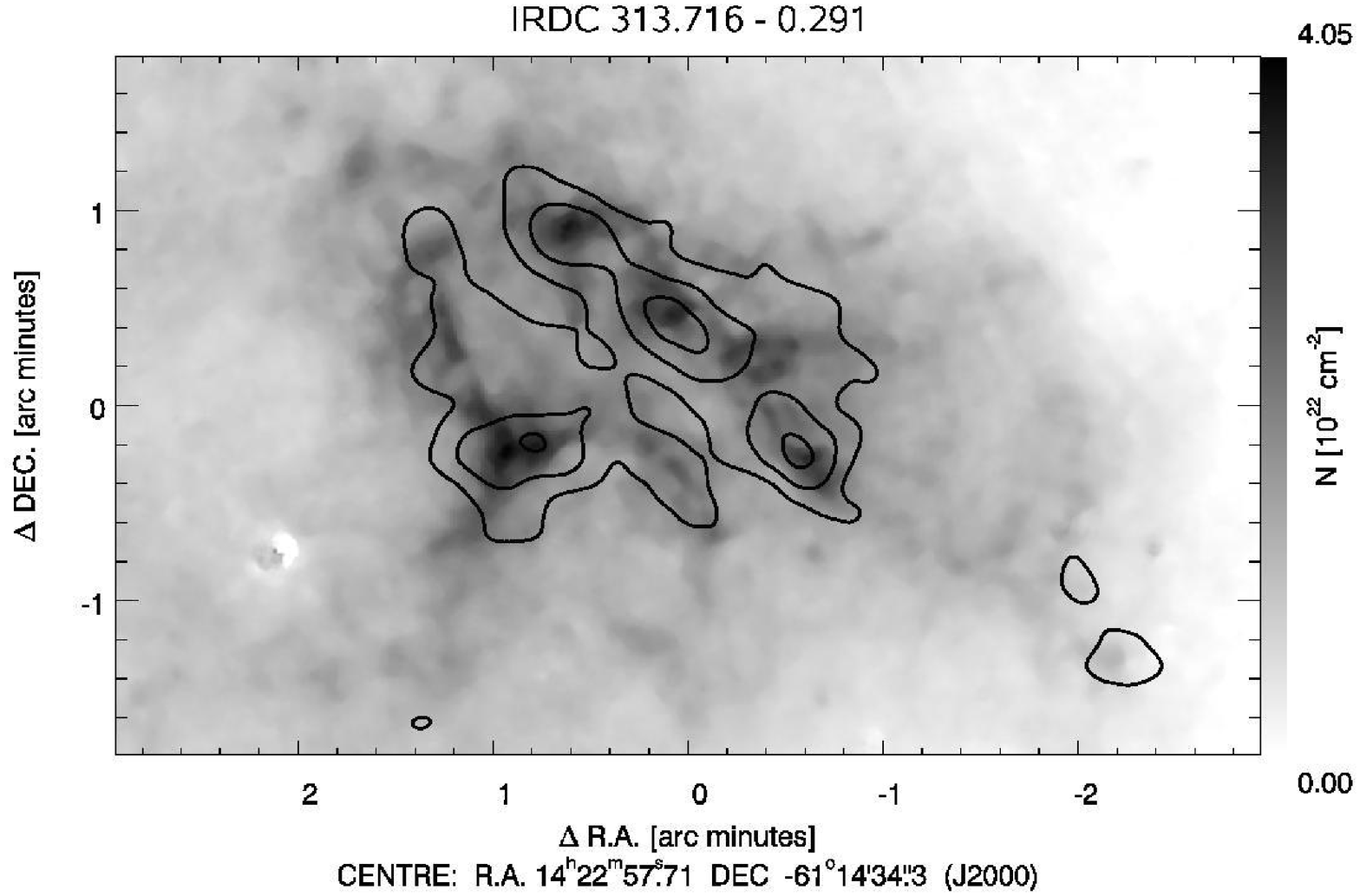}
   \includegraphics[width=7.8cm]{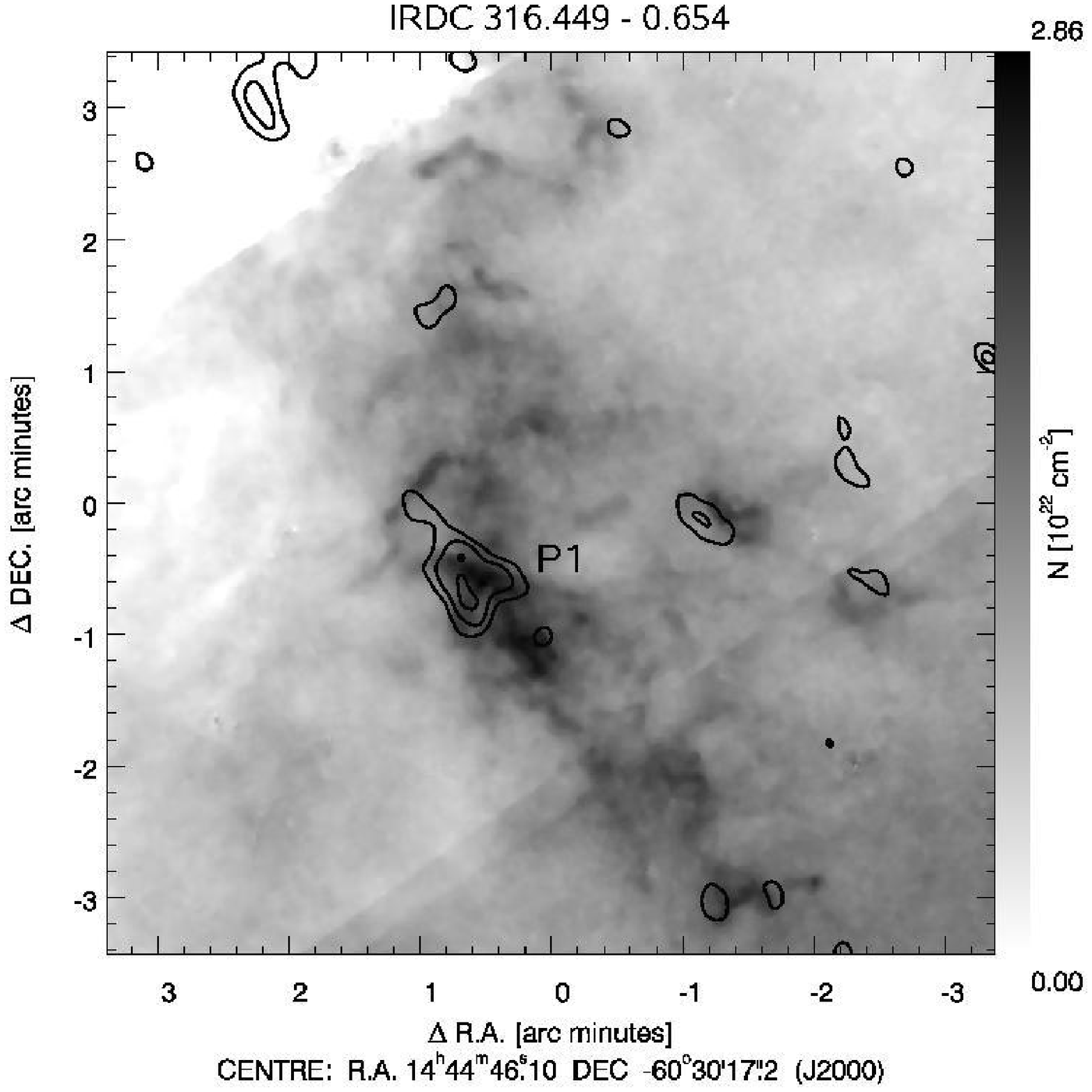}
   \includegraphics[width=7.8cm]{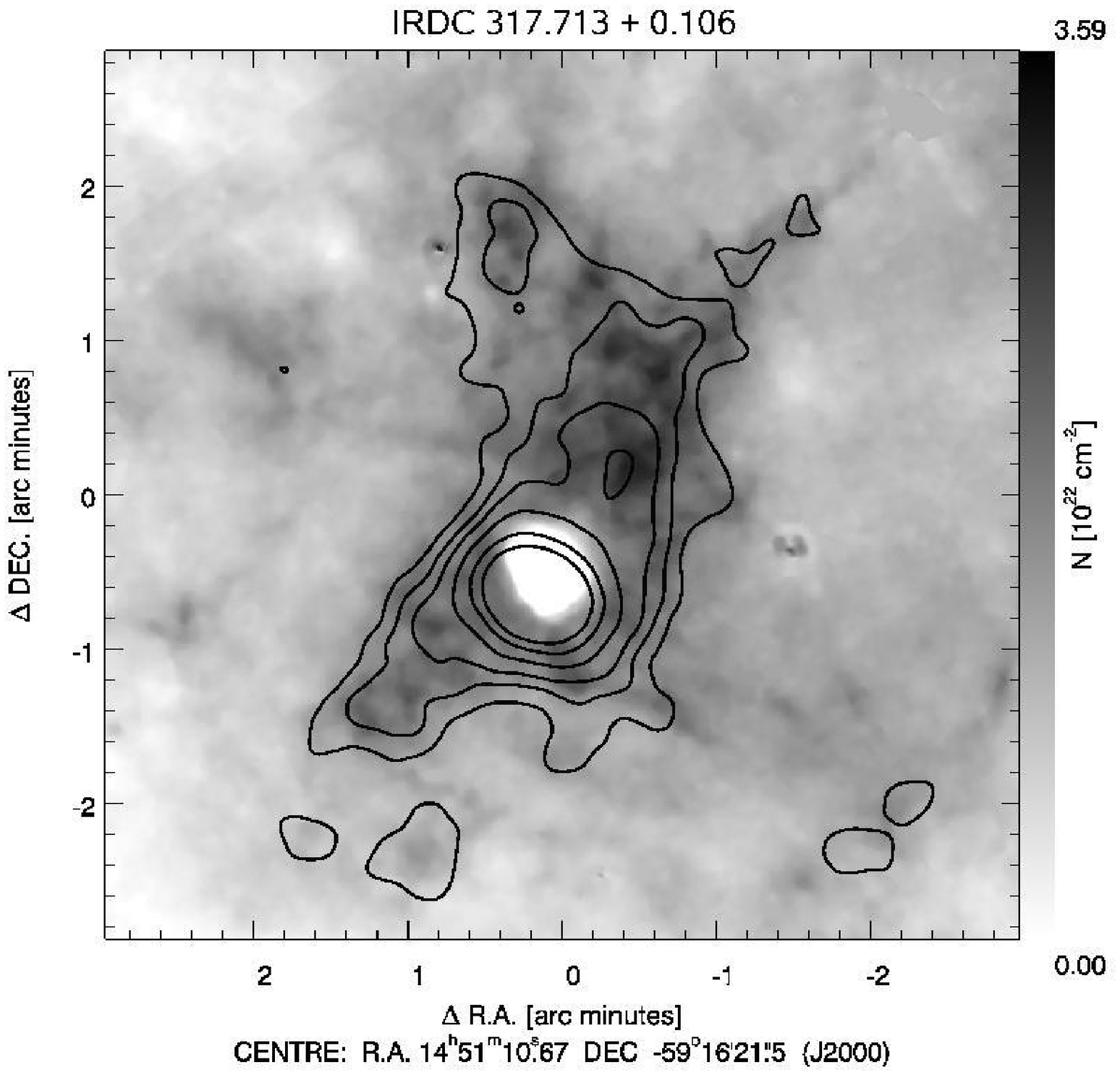}
   \includegraphics[width=7.8cm]{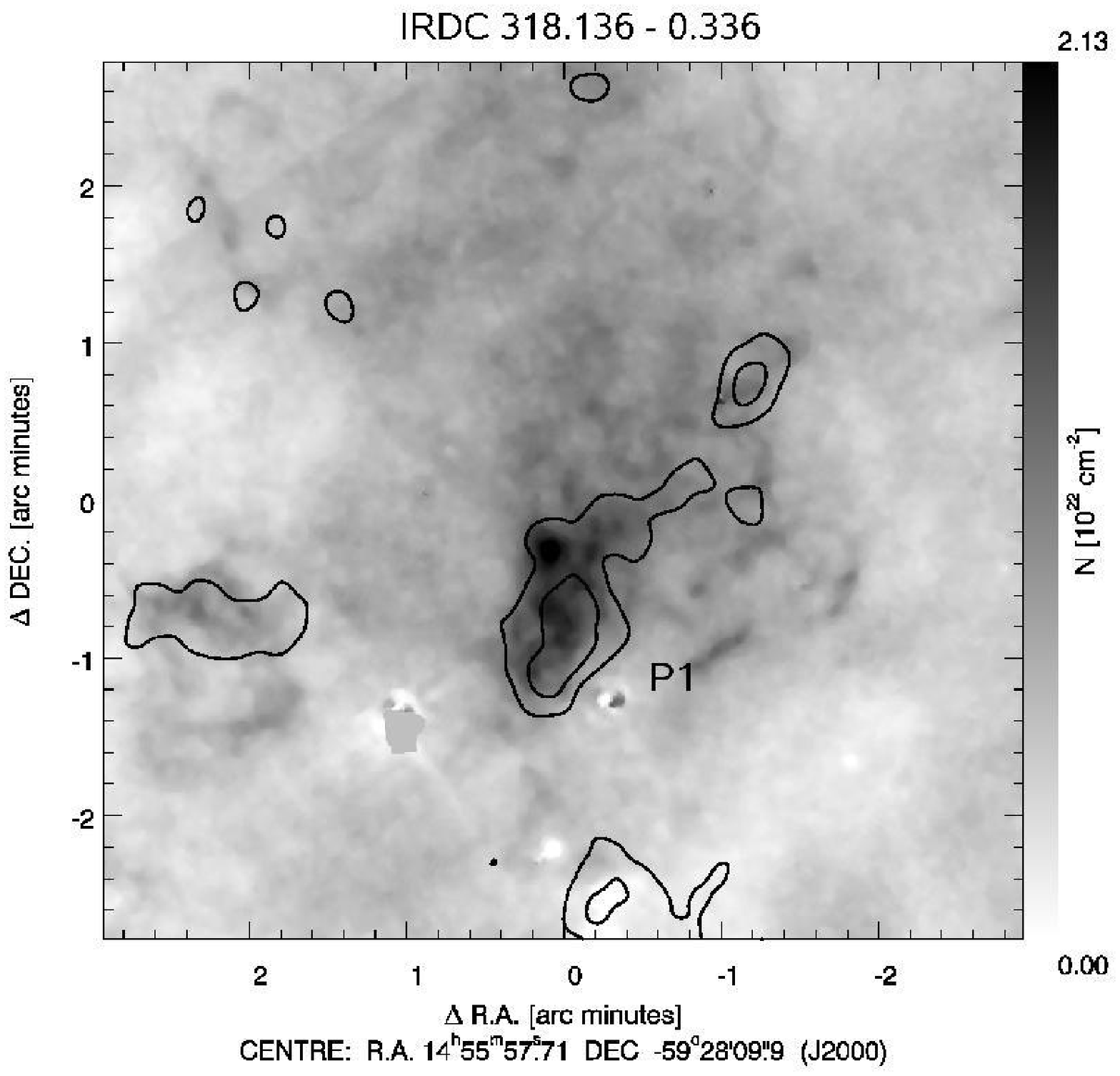}
   \includegraphics[width=7.8cm]{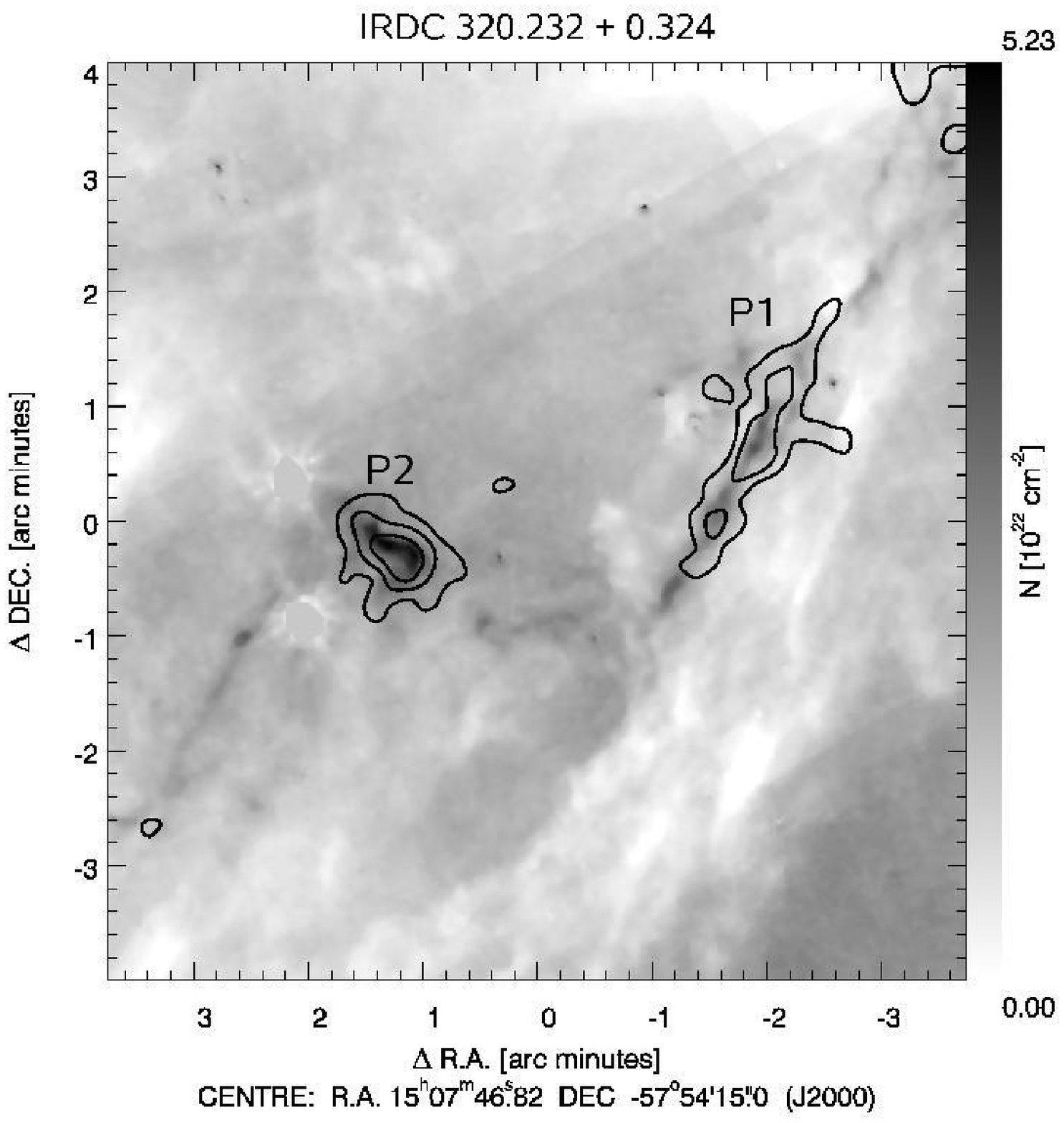}
   \includegraphics[width=8.3cm]{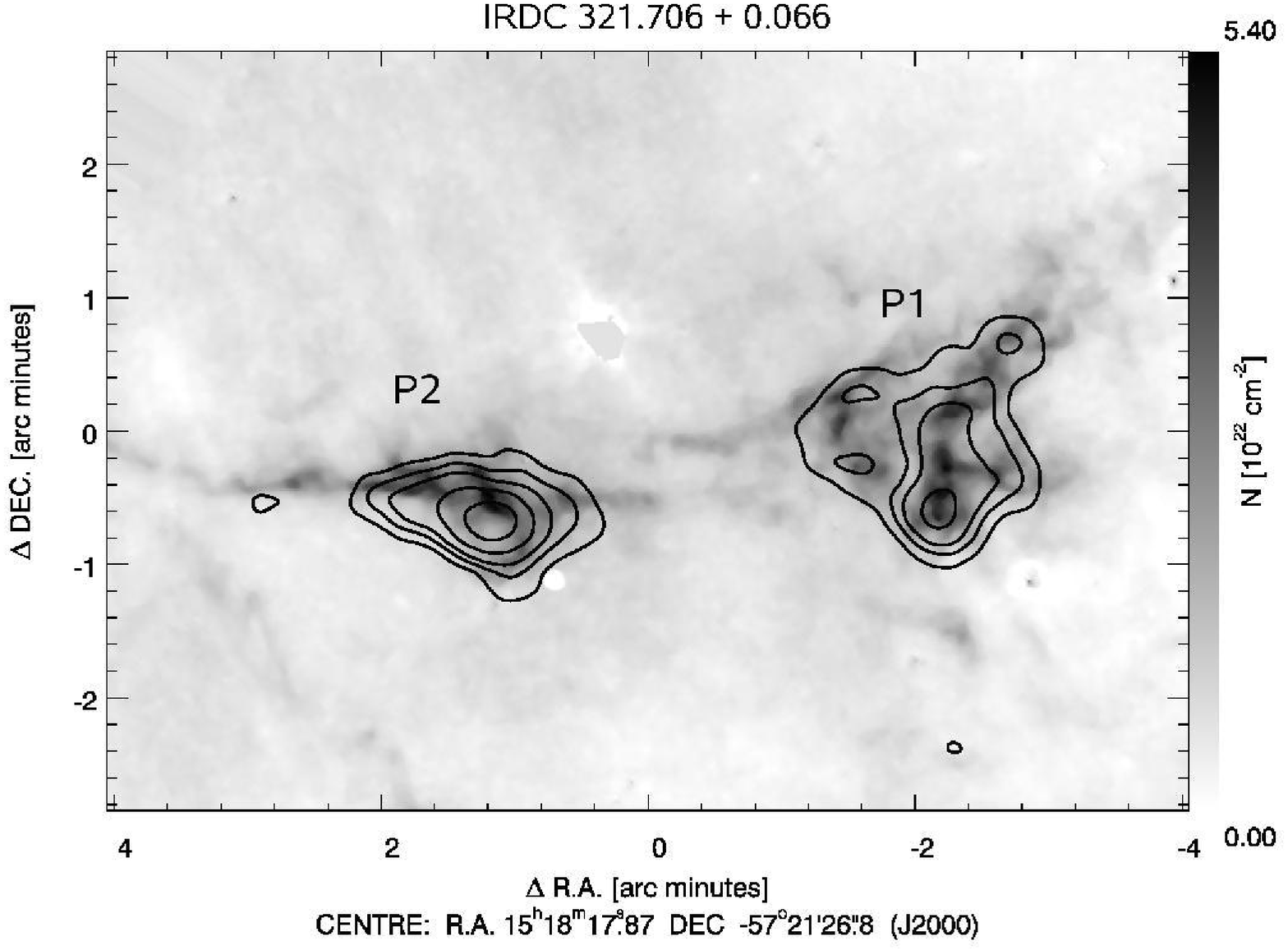}
      
      \caption{Column density maps derived from 8$\ \mu$m extinction overlaid with 1.2 mm 
   continuum emission as contours. The scaling is indicated in the bar to the right of each image.
   The contours are 60, 108, 156, 240, 360, 480 mJy beam$^{-1}$ in all cases except
   for IRDC 316.45-0.65 which is 84, 120, 156, 240, 360, 480 mJy beam$^{-1}$.
              }
         \label{Fig4}
   \end{figure*}  

\subsection{GLIMPSE 8~$\mu$m data}\label{Sect:GLIMPSE-Data}

The original selection of the IRDCs has been done still based on MSX images. 
In the meantime, the Spitzer satellite has succeeded MSX and provides much
higher spatial resolution and sensitivity.
GLIMPSE images for our regions with a pixel size of 0\farcs6 were retrieved from the 
NASA/IPAC Infrared Science Archive (IRSA) and re-mosaicked to cover the final field of our
1.2 mm maps of typically $15' \times 15'$ in order to get a picture of the IRDCs in relation 
to their closer and further vicinity (see Figs. ~\ref{Fig1}--\ref{Fig2}).
After bad pixel removal, a PSF photometry has been performed using the STARFINDER program
\citep{2000SPIE.4007..879D}. This allows to remove compact foreground objects, and 
thus maps of extended emission and absorption structures and finally column density maps 
could be extracted in a subsequent step. 

Dust masses for the IRDCs were computed assuming that they are in the foreground 
and shadow emission from behind. The optical depth $\tau$ is ideally the
logarithm of the ratio of two fluxes: (a) the flux from the emission background
$I_{\rm back}$ directly behind the IRDC, 
and (b) the actually measured remnant flux $I_{\rm IRDC}$ from the 
location of the IRDC.  Furthermore, superimposed on the IRDC is an emission 
contribution from foreground material, $I_{\rm fore}$, which has to be
subtracted. Since $I_{\rm back}$ cannot be directly estimated we need a measurable 
quantity $I_{\rm 0}$ that can be used as a proxy for $I_{\rm back}$. 
Then the optical depth is given by
   \begin{equation}\label{Equ-tau}
      \tau = \ln \left( \frac{I_{\rm 0}-I_{\rm fore}}{I_{\rm IRDC}-I_{\rm fore}} \right)
   \end{equation}
Following \citet{2008ASPC..387...50P} we assume that $I_{\rm fore}$ = $I_{\rm zl}$,
where $I_{\rm zl}$ is the zodiacal light, which is systematically calculated for
every Spitzer observation and available in the image header. For the quantity $I_0$ we
used the average emission level from a patch of MIR emission in the close vicinity of the 
actual cloud. These emission patches, typically around 1 square 
arcminute in size,  were chosen manually in order to capture the characteristic emission
level for the background approximation of the individual clouds and to exclude strong
compact emission sources. The mean and standard deviation of the 
emission levels within these defined regions have been computed in order to be used as 
$I_0$ in Equ.~\ref{Equ-tau}.
The standard deviation obtained here is propagated through the following steps and
is used to give formal errors for the derived masses and column densities, as listed
in Table \ref{table:1}. 

After optical depth determination, this quantity is converted into column densities and 
finally to masses by using the following equations:
   \begin{equation}
      N_{\rm H_2} =  1.086 \frac{\tau}{\sigma},
   \end{equation}
  
   \begin{equation}
      M =  {\rm m}_{\rm H_2} \ \  A \ \ N_{\rm H_2},
   \end{equation}
where $N_{\rm H_2}$ is the H$_2$ molecule column density, m$_{\rm H_2}$ is the mass of one 
hydrogen molecule, and $A$ is the area per pixel. $\sigma$ is the extinction cross 
section per hydrogen molecule. According to the adopted dust extinction model by \citet[][see below]{2001ApJ...548..296W}, 
$\sigma = 4.62 \times 10^{-23}\,{\rm cm}^2$ for the Spitzer/IRAC band 4 central wavelength of 7.87~$\mu$m.
   
The derived masses critically depend on the used extinction--column density
calibration, the distance to the targets, and the method to select the relevant
extinction regions.  
As  dust model we used the parametrisation of \citet{2001ApJ...548..296W} 
according to their model B with R$_{\rm V}$ = 5.5. This
particular model has been shown to be relevant for massive star--forming
regions, e.g., by \citet{2005ApJ...619..931I}. It differs from the common dust models 
\citep[e.g.,][]{1984ApJ...285...89D} in that it predicts higher extinction cross sections especially 
in the 4--8 micron wavelength region, a relevant point for the Spitzer extinction maps. We note 
in passing that such elevated cross sections are also predicted in connection with ice-coated
dust grains and especially if dust coagulation processes are involved 
\citep{1994A&A...291..943O}. The chosen dust model finally relates the
optical depth (and, equivalently, the extinction magnitude at the used
wavelength) to the equivalent column density of hydrogen molecules. 
The quadratic dependence of the masses on the distance is in the extinction map case the same as for 
masses derived from the 1.2 mm emission data.

\begin{table}
\caption{Comparison of the 1.2 mm emission and 8 $\mu$m absorption techniques for deriving masses and column
densities.} 
\label{table:2} 
\centering    
\begin{tabular}{l c c}        
\hline\hline
\noalign{\smallskip}
Properties &  1.2 mm &  8 $\mu$m  \\
\noalign{\smallskip}
\hline
\noalign{\smallskip}
Distance dependence$^{\mathrm{a}}$ &  + & +   \\
\noalign{\smallskip}
Resolution &  24$''$ & 3$''$    \\
\noalign{\smallskip}
Sensitive to lower column  &  - & +    \\
density filigree structure &    &     \\
\noalign{\smallskip}
Temperature dependence &  + & -    \\
\noalign{\smallskip}
Background and foreground &   not  &  necessary   \\
 estimation &  necessary &     \\
 \noalign{\smallskip}
Sensitive to column  &  + & -    \\
densities $\gg 10^{23}$\,cm$^{-2}$   &   &     \\
\noalign{\smallskip}
\hline                                  
\end{tabular}

\begin{list}{}{}
\item[$^{\mathrm{a}}$] Only for masses
\end{list}

\end{table}

\begin{table*}
\caption{Comparison of the different observational techniques for IRDC 18223-3.}  
\label{table:IRDC18223} 
\centering    
\begin{tabular}{l c c c c}        
\hline\hline
\noalign{\smallskip}
Data &  Resolution &  Column density& Optical depth& Contrast   \\
 & (arcsec)  & (10$^{22}$ cm $^{-2}$) & at 8\,$\mu$m & ($\rm I_{bg}/I_{IRDC}$)  \\
\noalign{\smallskip}
\hline
\noalign{\smallskip}
8~$\mu$m GLIMPSE &  3  & 2.3 & 0.95&2.5 \\
1.2 mm IRAM$^{\mathrm{a}}$&  11   & 5.9  & 2.4 &11 \\
3.2 mm PdBI$^{\mathrm{b}}$     &  5\farcs8 $\times$ 2\farcs4 & 45  & 19 & 1.8e+08 \\
1.3 mm SMA$^{\mathrm{c}}$ &  1\farcs3 $\times$ 1\farcs4 & 93 & 40  &1.8e+17 \\
\noalign{\smallskip}
\hline                                  
\end{tabular}

\begin{list}{}{}
\item[$^{\mathrm{a}}$]\citet{2002ApJ...566..945B}
\item[$^{\mathrm{b}}$]\citet{2005ApJ...634L.185B}
\item[$^{\mathrm{c}}$]Fallscheer, Beuther, et al.~2009 in preparation
\end{list}

\end{table*}

\subsection{Comparison of techniques}\label{Sect:Comparison}

Using the data from two different regions of the spectrum and two different techniques 
for estimating IRDC parameters, enables us to compare the results and to analyse advantages 
and disadvantages of both methods (see Table \ref{table:2}). The mid-IR data have an effective
resolution of around 3$''$, which is much higher than 24$''$ for our millimeter data. The estimated total masses 
and peak column densities have no dependence on the temperature. On the other hand,
when using 8~$\mu$m data to calculate the optical depth and then the column density in the cloud, 
we have to estimate the expected flux from the background indirectly. We assume the background 
flux as the difference between an average intensity around the cloud and the foreground 
intensity, but in reality matter hidden by the infrared dark cloud can be inhomogeneous and has 
either lower or higher intensity. 
However, the logarithm in Eq.~(\ref{Equ-tau}) to a certain degree
mitigates  uncertainties in the estimation of the intensities necessary for evaluating 
Eq.~(\ref{Equ-tau}). \\
\citet{2006ApJ...639..227S} have used an automated approach where they strongly
smooth the original (MSX) mid-infrared images and basically use these smoothed data as
approximation for the quantity $I_0$ (cf.~Eq.~\ref{Equ-tau}). We tested this approach for one of 
our clouds but found that unless very large smoothing kernels are used, the herewith derived 
intensity contrasts are systematically smaller than for the method we employed 
(Sect.~\ref{Sect:GLIMPSE-Data}). For very large smoothing kernels ($> 10'$) it is very difficult to
control the process. Continuum emission from strong extended emission sources in the surroundings
might be folded into the area of the IRDCs, depending on the individual circumstances. Therefore, we 
refrain from using this smoothing approach in order to have better control for the estimation of
the quantity $I_0$. \\
Uncertainties in the foreground level estimation also affect the resulting masses and column
densities. At the moment we take into account only the zodiacal light contribution, which typically accounts for 15\%--20\%
of the intensity toward the IRDCs. An extreme (hypothetical) case 
may be, that all remaining flux received from the IRDC, down to the noise level $I_{\rm noise}$ of the GLIMPSE images, 
is produced by another foreground emission contribution (e.g., the PDR surface of the cloud, glowing in PAH emission). 
After removing all this emission down to $I_{\rm noise}$, the resulting intensity contrast would of course be clearly
higher. However, considering typical levels for $I_{\rm noise}$ and $I_{\rm Back}$, our results for the column densities 
would rise just by a factor of around 3. This is again due to the alleviating effect of the logarithm in Eq.~(\ref{Equ-tau}), 
acting on the intensity contrast. 

   \begin{figure}
   \centering
     \includegraphics[width=8cm]{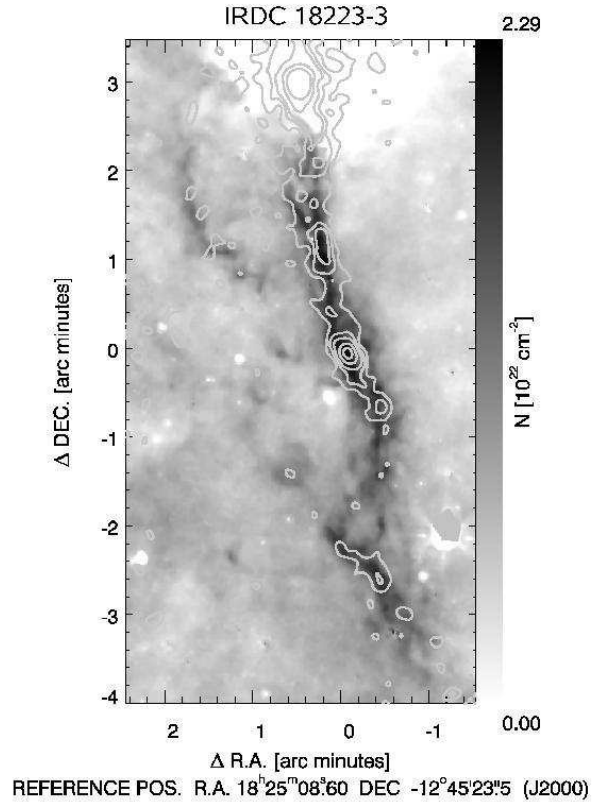}
     \caption{Column density map derived from an 8$\mu$m Spitzer/GLIMPSE image overlaid with 1.2 mm IRAM
   continuum emission as contours. The image scaling is indicated in the sidebar.
   The contours are 38,76,114,190,266 mJy beam$^{-1}$. The (0,0) position corresponds to the center
   of the IRDC 18223-3.}
     \label{Fig_I18223}
   \end{figure}

While for almost all our clouds peak column densities extracted from the GLIMPSE data are slightly higher than the 
values derived from the millimeter data, this difference does not correspond to the factor 8
in resolution. Therefore, we made a comparison between the peak column densities,
extracted from different observational data for the previously studied IRDC 18223-3
\citep{2002ApJ...566..945B,2005ApJ...634L.185B}, which is located 3.7 kpc from us and has a NH$_3$ rotation
temperature of $\sim$ 33 K \citep{2005ApJ...634L..57S}. 
For this particular cloud, 1.2 mm IRAM single-dish observations 
\citep{2002ApJ...566..945B}, 8 micron Spitzer/GLIMPSE data and interferometric data at 3.2 and  
1.3 mm data obtained with the PdBI and SMA \citep[][Fallscheer et al. 2009 in prep.]{2005ApJ...634L.185B} are available, 
with the spatial resolution ranging from 11 arcsec down to 1.4 arcsec (Table \ref{table:IRDC18223}). 
For the GLIMPSE data we obtained a column density distribution map according to the algorithm described in 
Sect.~\ref{Sect:GLIMPSE-Data}. Together with the 1.2 mm IRAM data as contours, this result is presented in Fig.~\ref{Fig_I18223}. 
The peak column densities for all the millimeter data were calculated using Eq.~(\ref{Equ:columndens}), adopting the
different beam sizes. Always a temperature of 33 K \citep{2005ApJ...634L..57S} has been used. The peak flux
density for the 3.2 mm 
PdBI data we took from the corresponding paper \citep{2005ApJ...634L.185B}, while we (re-)assessed the peak flux
densities for the 
IRAM\footnote{The 1.2 mm peak flux density we find in the IRAM 30-m data is clearly higher than reported in \citet{2002ApJ...566..945B}.
In that paper, the millimeter peaks had been fitted with Gaussians which occasionally underestimated the true peak flux densities.} 
and SMA data on the related FITS files, kindly provided by H.~Beuther and C.~Fallscheer. 
For the dust opacity per gram of dust, as for our SIMBA millimeter data, we have always used the same opacity model, 
appropriate for coagulated dust particles with thin ice mantles \citep[][opacity column 5 in their Table 1]{1994A&A...291..943O}.
As we can see in Table \ref{table:IRDC18223}, the peak column density derived from the high-resolution millimeter 
interferometry data is a factor of 40 higher than the one extracted with the mid-IR technique. A factor of a few 
between the mid-infrared and the millimeter single-dish results for the peak column densities  can be accounted for 
by using other dust opacities/extinction cross sections or more extreme assumptions on the MIR foreground 
contribution (see above). However, the large difference between the mid-infrared and the millimeter interferometry 
results indicates a principle limitation of the extinction method to distinguish high column density peaks. 
The realistically attainable intensity contrast at high optical depths sets this limit. In Table~\ref{table:IRDC18223}, we
list the optical depths and image contrasts at 8 micron that would correspond to the column densities derived from the millimeter
data. Obviously, realistic 8 micron images cannot provide such humongously high dynamic ranges necessary to derive column densities
similar to the interferometry results.

%__________________________________________________________________

\section{Results}

%__________________________________________________________________

\subsection{Morphology of the IRDCs}

The morphology of the clouds in our sample ranged from compact structures (IRDC 312.33-0.07
P1 and P2) to filaments (IRDC 309.13-0.15). They have sizes from 1$'$ (IRDC 013.84-0.49 
P2 and P3) to 4$'$ (IRDC 317.71+0.11), which roughly corresponds to 1--3.5 pc at 
the distance of these clouds. 
There is, in general, a good agreement between the morphologies 
of the 1.2 mm emission and the 8 $\mu$m extinction structures (see Fig. \ref{Fig3}-\ref{Fig4}). 
As a rule, dense areas of extinction material (IRDCs 013.84-0.49, 313.72-0.29 etc.) coincide 
with relatively bright compact sources at 1.2 mm. 
For some IRDC complexes, some \textit{emission} peaks at the millimeter wavelengths are associated 
with mid-IR \textit{emission} sources (IRDCs 309.13-0.15 P2, 309.34-0.64 P2 and 317.71+0.11) 
which indicates later evolutionary stages than those objects corresponding to 8 micron extinction.
Among all clouds there is one particular case - IRDC 310.39-0.28, where the millimeter 
emission  still peaks at the extinction maximum despite the existence of a bright 
MIR emission source nearby.\footnote{Based on the very similar LSR velocities for the MIR source and the extinction region,
derived from our recent molecular line observations, both objects are probably associated.}
The IRDCs 013.84-0.49, 312.33-0.07, 318.13-0.34, 320.23+0.32, and 321.71+0.07 present 
several separated sources at mm wavelengths, which coincide with dense 8 $\mu$m features.
IRDC 308.12-0.33 has an elongated shape and three sub-structures can be recognized,
one of them corresponds to the extinction maximum at mid-IR wavelengths. 
IRDC 309.13-0.15 shows two distinct 1.2 mm emission sources: the compact object (P1) coincides 
with the MIR extinction region, another one (P2) has an emission peak corresponding to the 
8\,$\mu$m emission and an elongated "tail". 
Extended millimeter emission, 2$' \times 4'$ in size, with at least four sub-structures 
is present in IRDC 313.72-0.29. 
For IRDC 316.45-0.66, we can recognise only one weak 1.2 mm source and diffuse extended MIR 
extinction structures.

Positions of known IRAS sources are marked with plus-signs on the 
Figs.~\ref{Fig1}--\ref{Fig2}, where the IRAC 8\,$\mu$m data are displayed as inverse grey-scale 
images, and contours present the 1.2 mm data. As a rule IRAS sources do not correspond to the extinction 
regions at 8\,$\mu$m. On the contrary, they agree with the locations where MIR emission coincides 
with millimeter emission peaks or just with the very bright MIR emission sources. In the case
of IRDC 317.71+0.11, the kinematic distance to one of the IRAS sources, located in the center of the 
8\,$\mu$m emission, can be derived using the CS (2--1) line velocities reported by 
\citet{1996A&AS..115...81B} toward this IRAS source. The $v_{\rm LSR}$ velocities of this CS measurement
and our HCO$^+$ data for the neighbouring IRDC are less than 0.5 km/s different.
The distance to the cloud and the IRAS source is then around 2.9 kpc. Hence, assuming that the dark 
cloud is related with this IRAS point source, the infrared luminosity of the compound (IR source + IRDC) 
is $\approx$ 10$^4$ L$_\odot$, using the IRAS approximation formula from \citet{1990A&A...227..542H}. 
This is the typical order of magnitude for pre-main-sequence OB stars.

\subsection{Masses and column densities}

Toward all 12 clouds millimeter continuum emission was detected and column
density maps were extracted from 8 $\mu$m images according to the algorithm 
described in Sect.~\ref{Sect:GLIMPSE-Data}. Figures \ref{Fig3}-\ref{Fig4} present these column density 
maps for every region, superimposed with the corresponding 1.2 mm contours. 
In Table \ref{table:1} the properties of the IRDCs are compiled: 
name, right ascension, declination, distance, peak flux density and integrated flux density,
mass and peak column density of the 1.2 mm sources, mass  
and peak column density of the extinction matter. For every cloud in the table, 
the first line corresponds to the total millimeter flux density as well as total masses of the emission and 
extinction matter. The following lines present data for the separate millimeter sub-clumps 
which are labeled with "P".

Where appropriate, we distinguished separate sub--clumps at 1.2 mm.
In the Fig. \ref{Fig3}-\ref{Fig4} and in
Table \ref{table:1} these separate emission sources are named with the designation 
"P" (e.g. P1, P2). For all of them we measured peak flux density and integrated flux density, derived 
masses and column densities according to Equ.~\ref{Equ:dustmass} - \ref{Equ:columndens}. 
In the case of IRDCs 309.13-0.15 (P2), 309.34-0.64 (P2) and 317.71+0.11 (P1) 
parts which coincide with strong 8 $\mu$m emission features were not taken into account for measuring 
1.2 mm flux densities and hence for estimating masses and column densities. 
The typical range of 
masses of the separate millimeter sources with $T$ = 20 K is 50-1000 $\rm M_\odot$, 
and the column densities range between 0.9 and 4.6 $\times  10^{22}$ cm$^{-2}$.

For calculating total masses of the extinction material in individual clumps we took regions 
above 3 sigma into account, where sigma is in this case the standard deviation in the full extinction 
maps.
The total masses of the IRDCs for extinction matter were found to range from 300 
to 1700 $\rm M_\odot$ and the derived peak column densities correspond to values from 
2.1 to 5.4 $\times  10^{22}$ cm$^{-2}$.

\begin{landscape}
\begin{table}

\caption{Properties of the IRDCs. For every cloud  
the first line corresponds to the total mass, the following to the masses 
of separate millimeter sources.} %title of Table
\label{table:1}      % is used to refer this table in the text
%\centering                          % used for centering table
\begin{tabular}{r c c c c c c c c c c c c c c}        
\hline\hline             % inserts double horizontal lines
\noalign{\smallskip}
Name & R.A. & Decl. & D & Peak Flux density & Integrated Flux density & Mass 1.2mm & N (1.2 mm)$^1$ &
N$_0$ (1.2 mm)$^2$ & Mass (8 $\mu$m) & N (8 $\mu$m)\\ 
& (J2000.0) & (J2000.0) & (kpc)& (mJy) & (Jy) & ($\rm M_\odot$)&
(10$^{22}$ cm $^{-2}$) & (10$^{22}$ cm $^{-2}$) &($\rm M_\odot$) & (10$^{22}$ cm $^{-2}$) \\   

\noalign{\smallskip}   
\hline                        % inserts single horizontal line
\noalign{\smallskip}
IRDC 308.12$-$0.33 & 13 37 01.2 &$-$62 44 40 & 4.32& 225 & 1.88 & 580 & 1.7  &60.3& 520$^{+40}_{-30}$  & 2.8$^{+0.16}_{-0.16}$ \\[1.5mm]
IRDC 309.13$-$0.15 & 13 45 17.1 &$-$62 21 57 & 3.92& 330$^3$ & 4.39$^3$ & 1150$^3$& 2.6$^3$  & &1750$^{+370}_{-310}$& 3.0$^{+0.13}_{-0.13}$ \\
P1 &  &  &  &  162 &  1.37 & 360& 1.3 & 41.9 & & \\	    
P2 &  &  &  &  124 &  0.96 & 250& 0.9 \\[1.5mm] 
IRDC 309.34$-$0.64 & 13 48 39.8 &$-$62 47 26 &3.46 & 504$^3$ & 3.69$^3$ & 750$^3$ & 3.9$^3$  & & 750$^{+170}_{-150}$& 3.1$^{+0.24}_{-0.24}$\\
P1 &  &  &  &  216 &  0.94 & 190& 1.7 & 48.5 & &\\
P2 &  &  &  &  155 &  0.81 & 170& 1.3 \\[1.5mm]
IRDC 310.39$-$0.28 & 13 56 01.7 &$-$62 14 27 & 4.93& 594 & 2.48 & 1029& 4.6  &186.1& 1320$^{+60}_{-50}$ & 2.6$^{+0.16}_{-0.16}$ \\[1.5mm]
IRDC 312.33$-$0.07 & 14 11 56.8 &$-$61 29 25 & 4.05& 288 & 0.77 & 210 & 2.3  &76.6& 290$^{+100}_{-60}$ & 2.7$^{+0.33}_{-0.16}$ \\
P1 &  &  &  &  288 &  0.46 & 130& 2.3 \\   
P2 &  &  &  &	92 &  0.31 & 90 & 0.7 \\[1.5mm]
IRDC 313.72$-$0.29 & 14 23 05.4 &$-$61 14 48 & 3.33& 172 & 1.98 &  370& 1.3  &35.7& 700$^{+90}_{-70}$  & 4.1$^{+0.16}_{-0.16}$ \\[1.5mm]
IRDC 316.45$-$0.66 & 14 44 50.4 &$-$60 30 54 & 3.01& 159 & 0.96 &  150& 1.3  &32.3& 430$^{+80}_{-70}$  & 2.9$^{+0.16}_{-0.16}$ \\
P1 &  &  &  &  159 &  0.58 & 90 & 1.3 \\[1.5mm]  
IRDC 317.71$+$0.11 & 14 51 07.5 &$-$59 16 11 & 2.9 & 954$^3$ & 8.00$^3$ & 1150$^3$ & 7.5$^3$  &       &1320$^{+150}_{-150}$& 3.6$^{+0.15}_{-0.15}$ \\
P1 &  &  &  &  250 &  1.98 & 280& 2.0 & 47.9 & & \\[1.5mm]
IRDC 318.13$-$0.34 & 14 55 58.4 &$-$59 28 31 & 2.96& 142 & 2.48 &  370& 1.1  &26.9& 680$^{+160}_{-140}$& 2.1$^{+0.16}_{-0.16}$ \\
P1 &  &  &  &  142 &  1.02 & 150& 1.1 \\  
%P2 &  &  &  &  120 &  1.07 & 160& 0.9 \\[1.5mm]
IRDC 320.23$+$0.32 & 15 07 56.7 &$-$57 54 27 & 1.97& 186 & 1.68 &  110& 1.5  &24.6& 600$^{+320}_{-250}$& 5.2$^{+0.16}_{-0.16}$ \\
P1 &  &  &  &  142 &  1.00 & 70 & 1.1 \\  
P2 &  &  &  &  186 &  0.68 & 50 & 1.5 \\[1.5mm]
IRDC 321.71$+$0.07 & 15 18 26.7 &$-$57 21 56 & 2.14& 408 & 3.12 &  240& 3.2  &56.9& 460 $^{+80}_{-80}$ & 5.4$^{+0.16}_{-0.23}$ \\
P1 &  &  &  &  276 &  1.69 & 130& 2.2 \\  
P2 &  &  &  &  408 &  1.43 & 110& 3.2 \\[1.5mm]
IRDC 013.84$-$0.49 & 18 17 21.2 &$-$17 09 23 & 2.66& 438 & 9.38 & 1130& 3.5  &77.0&1150$^{+130}_{-120}$& 3.3$^{+0.13}_{-0.13}$\\
P1 &  &  &  &  214 &  2.97 & 360& 1.6 \\
P2 &  &  &  &  163 &  1.12 & 130& 1.3 \\
P3 &  &  &  &  192 &  1.45 & 170& 1.5 \\  
P4 &  &  &  &  438 &  3.54 & 430& 3.5 \\	
\noalign{\smallskip}   
\hline                                   %inserts single line
\end{tabular}

\end{table}

$^1$ Peak column density per beam

$^2$ Extrapolated column density, obtained by applying the correction factor explained in Sect.~\ref{Sect:high-comp}

$^3$ This mm peak corresponds to a mid-infrared bright source near to the IRDC.

\end{landscape}

\subsection{Comparison with results for other cores}\label{Sect:comp2}

In this chapter
we will compare our results with previously obtained
characteristics for low- and high-mass pre-stellar cores.

\subsubsection{Comparison with low-mass cores}\label{Sect:low-comp}
\begin{figure}
   \centering
   \includegraphics[width=6.0cm,angle=90]{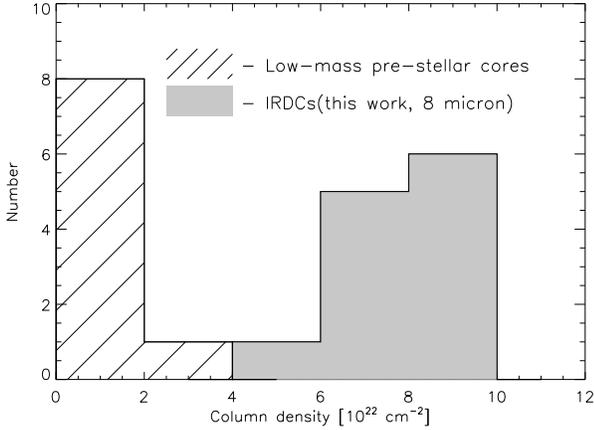} 
      \caption{Histogram showing the principle distribution of column densities for a collection of
               low-mass cores from \citet{2000A&A...361..555B} (after taking into account consideration
	       (a) from Sect.~\ref{Sect:low-comp}) and for the IRDCs presented here (after taking into 
	       account consideration (b) from Sect.~\ref{Sect:low-comp}). The Mann-Whitney-U one-tailed test confirms the 
               clear separation of the two distributions even when the different spatial resolutions (consideration (b))
	       are not taken into account (Sect.~\ref{Sect:low-comp}).
              }
         \label{Fig:low-high-comp}
   \end{figure}

During recent years, extinction mapping has mainly been applied  
along the lines of the near-infrared colour excess method \citep[e.g.,][]{2001A&A...377.1023L, 2008arXiv0809.3383L} or
classical star count techniques in the visible or near-infrared 
\citep[e.g.,][]{2005PASJ...57S...1D,2005A&A...432L..67F}. Such approaches are most powerful towards 
medium-extinction regions. In contrast, the extinction map method used in the current paper exploits the 
extinction of mid-infrared extended emission and hence does not rely on the identification of stellar background sources.
This method can peak into cores having column densities of up to $10^{23}$ cm$^{-2}$
%(corresponding to $A_{\rm V} \approx 60$ mag, cf. Sect.~\ref{Sect:Comparison})
%\footnote{We will discuss the limitations of this technique in Sect.~\ref{Sect:Comparison}.}. 
It has been used previously for the analysis of low-mass
starless cores based on ISOCAM data \citep{2000A&A...361..555B} which gives us the opportunity to compare 
our results. To make a fair comparison, two effects have to be considered. \textbf{(a)} \citet{2000A&A...361..555B}
used the standard \citet{1984ApJ...285...89D} extinction cross sections. These are more than a factor of
3 smaller than the \citet{2001ApJ...548..296W} values we use. Therefore, we recomputed the peak column 
densities reported in \citet{2000A&A...361..555B} for the low-mass cores by adopting the 
\citet{2001ApJ...548..296W} dust extinction model. \textbf{(b)} For the low-mass 
cores, which typically reside at distances of less than 300 pc, the linear spatial resolution is much better
than in the case of IRDCs with their distances of 2 - 5 kpc. Therefore, the actual column density peaks
are much better resolved in the low-mass case. To assess the effect of poor spatial resolution on our
derived peak column densities, we used a synthetic column density map derived in \citet{2005A&A...434..167S} 
for one of the low-mass cores of \citet{2000A&A...361..555B}, namely Rho Oph D ($d$ = 160 pc). We convolved 
this map with kernels appropriate to emulate the much coarser linear resolution toward our IRDC targets and
computed the ratio of the unsmoothed to the smoothed peak column density value. These factors (typically in 
the range from 1.5 to 3.5) have been multiplied to the peak column densities we have originally derived from 
the IRDC extinction maps. \\
A comparison of the column densities for low-mass cores and our IRDCs, after taking into account the just 
mentioned considerations (a) and (b), is shown in Fig.~\ref{Fig:low-high-comp}. Here 
we can see a clear trend for high-mass cores to attain higher column densities than the low-mass objects.
This qualitative difference shows that IRDCs are not just far away Taurus-like clouds, but a distinct type 
of clouds with the potential to form a distinct type of stars (see below and Sect.~\ref{Sect:Theo-Comp}).

It might be of concern that in step (b) we applied a correction factor which increases the column densities 
with increasing kinematic distance.
To statistically fortify our statement that Infrared Dark Clouds present ``stochastically larger'' 
values for column densities than low-mass pre-stellar cores we used the (Wilcoxon)-Mann-Whitney U one-tailed 
test \citep[e.g.,][]{2003psa..book.....W}. It is a non-parametric test for assessing whether two samples of 
observations come from the same distribution or not. It works well on a small number of observations in one 
sample. The null hypothesis is that the two samples are drawn from a single population. For the test we 
used low-mass column density values  from \citet{2000A&A...361..555B} adopted to the  
\citet{2001ApJ...548..296W} dust extinction model and column densities for IRDCs \textit{without} applying the 
linear spatial resolution correction (b) mentioned above.

For both populations we computed the  Rank Sum within the nonparametric Mann-Whitney statistic, 
usually called U (108 vs 0). The distribution of the U-statistic under the null hypothesis is known and can be found in 
special tables \citep[e.g.,][]{1988psa..book.....S}. Then, we estimated the probability that the values for low-mass 
objects and IRDCs came from the same distribution. The obtained probability lower than 0.005 rejects the null 
hypothesis and shows that our two samples come from different distributions.

\subsubsection{Comparison with high-mass cores}\label{Sect:high-comp}
The high-mass clumps we want to compare have all been observed at 1.2 mm, comprising the following samples: 
high-mass starless core candidates (HMSCs) \citep{2005ApJ...634L..57S,2002ApJ...566..945B}, 
Infrared Dark Clouds from \citet{2006ApJ...641..389R}, and results presented in this paper for our 1.2 mm data.
To (re-)calculate peak column densities for the first two sets of data we used  Eq.~(\ref{Equ:columndens}) 
as well. In the case of HMSCs, peak flux densities were taken from \citet{2002ApJ...566..945B}, distance and temperature 
estimates for all these objects come from \citet{2005ApJ...634L..57S}. 
In  \citet{2006ApJ...641..389R}, for all 38 IRDCs peak flux densities and distances are presented. For the calculation,
we assume again 20 K for the temperature, as dust opacity $\varkappa_v$ and gas-to-dust mass ratio we adopt the 
values 1.0 cm$^2$ g$^{-1}$ and 100, respectively. 
Hence, we use here the same parameters as for the analysis of our 1.2 mm data, except for the different 
beam size, equal to (11$''$)$^2$ in steradians, adapted to the IRAM 30-m telescope. 

Still, the measured data are affected by the convolution with the observational beam, which results in different 
linear smoothing scales for objects at different distances.
In order to eliminate this smoothing effect, we attempted to extrapolate the measured peak column densities per beam back
to the true values, assuming an (analytic) column density profile. % $\sim$ r$^{-1}$.
But what is an appropriate choice for such a power law? A first idea is to use the data at different beam sizes we have for IRDC
18223-3 (Sect.~\ref{Sect:Comparison}). However, it turns out that these data sets give no unique trend for such a power law. The 
comparison between the IRAM 30-m data and the SMA data speak for a relatively steep power law of $N  \sim$ r$^{-1.75}$, and the comparison 
with the PdBI data would even indicate a much steeper power law index. The fact that these two data sets do not result in a similar
power law may be related to the different degree these observations are able to recover extended emission. Furthermore, at the 
high spatial resolution of the SMA observations, we begin to see a dense rotation structure (Fallscheer et al. 2009, in prep.) that is 
distinct from the rest of the surrounding clump. 
However, we also compared the column density values for a few other IRDCs, observed both at single-dish and interferometric resolution 
\citep{2006ApJ...641..389R,2008arXiv0808.2973R}. Again, no clear trend for a certain power law range is obvious.
According to \citet{2003ApJ...588L..37J}, for the filament structure of the famous IRDC G11.11-0.02 the column density profile 
should be very steep and may reach $\sim$ r$^{-3}$. Such a significant difference to commonly observed values might be an 
additional feature characterizing massive star forming clumps, especially if filamentary structures are involved.
On the other hand, \citet{2000A&A...361..555B} showed that N $\sim$ r$^{-1}$ can be a reasonable choice for
lower-mass starless cores, and also the single dish mapping of young massive clumps by \citet{2002ApJ...566..945B} 
resulted in less steep power laws for the column density quite close to the low-mass core results. 
Hence, by assuming N $\sim$ r$^{-1}$  for the extrapolation to the true column densities will provide us with 
a kind of robust lower limit.

To recalculate all column density values, an artificial column density distribution map was created (by just assuming
an analytical N $\sim$ r$^{-1}$ profile) with the peak in 
the centre of the array normalised to 1.0 and the assumption that 1 pixel corresponds to 2000 AU (a typical size scale for
fragmentation). 
Then, for every source from the three high-mass samples mentioned above, this column density distribution map has been 
convolved with a smoothing Gaussian kernel emulating the effect of observing at mm wavelengths with single-dish telescopes 
(11--24$''$ beam). The kernel size (FWHM) is computed by taking the ratio of the effective linear resolution (in AU) achieved 
within the observed mm maps and the 2000 AU pixel size from the artificial map.
After convolving the artificial map with such a Gaussian kernel, the peak column densities in these 
synthetic maps are smaller than 1. The correction factor for every source is then the ratio of the original 
peak column density value (1.0) to the peak column density value after beam convolution. These factors are then applied
to the column density values derived according to the Eq.~(\ref{Equ:columndens}), which then approximates the true 
peak column density values. These extrapolated values for the new regions reported here are also given in column 9 of 
Table \ref{table:1}. 

\begin{figure}
   \centering
   \includegraphics[width=6.0cm,angle=90]{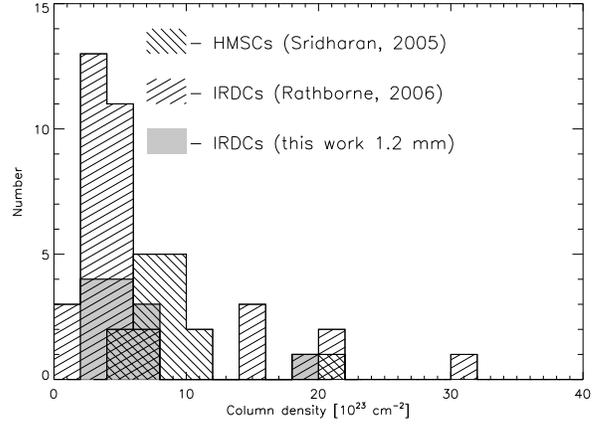} 
      \caption{Distribution of extrapolated column densities for the HMSCs from \citet{2005ApJ...634L..57S}, 
      the IRDCs from \citet{2006ApJ...641..389R}, and the IRDCs from the present paper. Note that we have applied the
      corrections and extrapolations mentioned in Sect.~\ref{Sect:high-comp} for all these high-mass cores.
              }
         \label{Fig5}
   \end{figure}

Figure \ref{Fig5} summarizes the distribution of the extrapolated peak column densities we have finally obtained.
The distribution of the true peak column densities indicates a similar order of magnitude for previously obtained 
values for the two other samples of massive star forming regions and our new SIMBA results.
Most of the clouds have (extrapolated) column densities in the range from 1.0 to 12.0 $\times  10^{23}$ cm$^{-2}$ with
few exceptions reaching 20.0 - 30.0 $\times  10^{23}$ cm$^{-2}$.

\subsection{Comparison with theoretical models.}\label{Sect:Theo-Comp}

Infrared Dark Clouds, located several kpc away from us, have large masses, volume
and column densities, but ``high'' values alone do not guarantee that they are really the
progenitors of massive stars, and we need additional criteria to estimate this 
possibility.
According to \citet{2008Natur.451.1082K} only
clouds  with column densities of at least 1 g cm$^{-2}$, which 
corresponds to 3 $\times  10^{23}$ cm$^{-2}$ in our units, can form massive stars. 
As we can see in Table \ref{table:1}, the direct transformation of observational results 
for IRDCs from our list indicates peak column densities one order of magnitude lower than this limit. 
On the other hand, as it was shown in Sect.~\ref{Sect:comp2}, that the directly observed peak column densities 
are still affected by the convolution with the beam of the millimeter observations. After taking this into
account, the derived values for the extrapolated column density rise by a factor of 
10 or higher (see Fig.~\ref{Fig5}). Thus, we can reach the 3 $\times  10^{23}$ cm$^{-2}$  
threshold in the case of the infrared dark clouds.

Since we have not only compact sources, but also  very filamentary structures like IRDC 320.23+0.32 P1, this 
raises the question, whether it is possible to form massive stars in such structures.
\citet{2008ASPC..387..216B} showed that filaments play a dominant role in controlling the physics, 
accretion rate, and angular momentum of the much smaller-scale accretion disk that forms within 
such collapsing structures. Large-scale filamentary flows sustain a very high accretion rate 
($\dot{M} \sim$ 10$^{-2} \ \rm M_\odot \ $yr$^{-1}$) due to the supersonic
gas flow onto the protostellar disk. These rates are 10$^3$ times larger than predicted by the collapse
of the singular isothermal spheres and exeed the accretion rates necessary to squeeze the radiation
field of the newly born massive star.
Thus, for almost all our clouds we still see the potential to form massive stars.

%______________________________________________________________

\section{Conclusions}

%   \begin{enumerate}

In this paper, we discuss our progress in understanding properties of Infrared Dark Clouds. A set
of 12 clouds located in the southern hemisphere has been selected from the MSX 8.3 micron images. 
For these clouds 1.2 mm maps were obtained with the SIMBA bolometer array at the SEST telescope.
GLIMPSE mid-infrared images for these regions were retrieved from the Spitzer Archive. 

The new sources comprise a variety of IRDC morphologies, from compact cores to filamentary
shaped ones, from infrared quiet examples (no Spitzer 8 $\mu$m emission sources) 
to more active ones. 
As a rule, our sample shows a good agreement between the morphologies of 1.2 mm emission 
and 8 $\mu$m extinction features.
The total masses of the IRDCs were found to range from 150 to 1150 $\rm M_\odot$ 
(emission data) and from 300 to 1750 $\rm M_\odot$ (extinction data). We derived peak column 
densities between 0.9 and 4.6 $\times  10^{22}$ cm$^{-2}$ (emission data) and 
2.1 and 5.4 $\times  10^{22}$ cm$^{-2}$ (extinction data).

Since the MIR extinction method has been used previously for the analysis of low-mass
starless cores we check how our findings relate to the published results. To make a fair comparison, 
we used the same dust model and the same spatial resolution in both cases. It is shown, that there is a
clear trend for the high mass cores to attain higher column densities than the low-mass objects.
This qualitative difference means that most IRDCs are not just far away Taurus-like clouds, 
but a distinct type of clouds with the potential to form a distinct type of stars.
A simple statistical analysis (the Mann-Whitney-U one-tailed test) confirms this statement also
when the different spatial resolutions are not taken into account and thus the IRDC column densities are
underestimated.

Using the data from two different regions of the spectrum and applying two different techniques 
for estimating IRDC parameters enables us to compare these two methods.
On the one hand, the extinction technique has some advantages over the millimeter technique. 
It is a "cheap" method since GLIMPSE at 8 $\mu$m has covered large parts of the Galactic plane 
in the 4th and 1st quadrant. The GLIMPSE data 
have a better spatial resolution than millimeter single-dish data, which reveals the often filigree 
substructures of the clouds. Furthermore, this method does \textit{not} depend on assumptions 
for the temperature of the IRDCs for estimating masses and column densities.
However, our comparison shows, that the extinction method has a
principle limitation to distinguish very high image contrasts and hence to find high column density peaks 
 $\gg 10^{23}$ cm$^{-2}$. Hence with the extinction method we can give only a lower limit to 
the column density values, inspite of high resolution. The limit is around $A_{\rm V} = 75$ mag when 
applying the \citet{2001ApJ...548..296W} $R_{\rm V} = 5.5$ B extinction law 
\citep[corresponding to roughly 200 mag when following the common $R_{\rm V} = 3.1$ extinctionlaw reviewed in][]{1990ARA&A..28...37M}. 
High-spatial resolution (sub-)millimeter observations are hence crucial to assess even higher column 
density ranges and to reveal the actual column density maxima.

To compare column densities extracted with the emission method with previously obtained values for 
IRDCs and HMPOs, we extrapolated them back to the true peak column densities by assuming a column density profile 
$\sim$ r$^{-1}$, thus, mitigating the spatial resolution differences within the different samples. 
The distribution of the true peak column densities indicates a similar order of 
magnitude for our new SIMBA results and the two other samples of massive star forming regions.
Moreover, the true peak column densities exceed the theoretical limit of 3 $\times  10^{23}$ cm$^{-2}$ (or 1 g cm$^{-2}$),
recently put forth to distinguish potentially high-mass star-forming clouds.

Thus, extracted values for masses and column densities both for emission and extinction matter 
show a clear difference between IRDCs and known low-mass pre-stellar cores, and confirm 
our assumptions that Infrared Dark Clouds can present the earliest stages of high-mass 
star formation. 

%   \end{enumerate}

\begin{acknowledgements}

We thank J\"urgen Steinacker for discussions and for providing us with the sythetic column
density map of Rho Oph D. We are indebted to Henrik Beuther and Cassandra Fallscheer for discussions 
and for making available the data for IRDC 18223-3 to us in electronic form. We wish to thank 
Maxim Voronkov for help with observations with the Australian MOPRA telescope. 
This research has made use of the NASA/ IPAC Infrared Science Archive, which is operated by 
the Jet Propulsion Laboratory, California Institute of Technology, under contract with the 
National Aeronautics and Space Administration. NASA's Astrophysics Data System was used to assess
the literature given in the references.

\end{acknowledgements}

\bibliographystyle{aa}
\bibliography{IRDCs_clean}

\begin{thebibliography}{40}
\expandafter\ifx\csname natexlab\endcsname\relax\def\natexlab#1{#1}\fi

\bibitem[{{Bacmann} {et~al.}(2000){Bacmann}, {Andr{\'e}}, {Puget}, {Abergel},
  {Bontemps}, \& {Ward-Thompson}}]{2000A&A...361..555B}
{Bacmann}, A., {Andr{\'e}}, P., {Puget}, J.-L., {et~al.} 2000, \aap, 361, 555

\bibitem[{{Banerjee} \& {Pudritz}(2008)}]{2008ASPC..387..216B}
{Banerjee}, R. \& {Pudritz}, R.~E. 2008, in Astronomical Society of the Pacific
  Conference Series, Vol. 387, Astronomical Society of the Pacific Conference
  Series, ed. H.~{Beuther}, H.~{Linz}, \& T.~{Henning}, 216--+

\bibitem[{{Benjamin} {et~al.}(2003){Benjamin}, {Churchwell}, {Babler}, {Bania},
  {Clemens}, {Cohen}, {Dickey}, {Indebetouw}, {Jackson}, {Kobulnicky},
  {Lazarian}, {Marston}, {Mathis}, {Meade}, {Seager}, {Stolovy}, {Watson},
  {Whitney}, {Wolff}, \& {Wolfire}}]{2003PASP..115..953B}
{Benjamin}, R.~A., {Churchwell}, E., {Babler}, B.~L., {et~al.} 2003, \pasp,
  115, 953

\bibitem[{{Beuther} {et~al.}(2007){Beuther}, {Churchwell}, {McKee}, \&
  {Tan}}]{2007prpl.conf..165B}
{Beuther}, H., {Churchwell}, E.~B., {McKee}, C.~F., \& {Tan}, J.~C. 2007, in
  Protostars and Planets V, ed. B.~{Reipurth}, D.~{Jewitt}, \& K.~{Keil},
  165--180

\bibitem[{{Beuther} {et~al.}(2002){Beuther}, {Schilke}, {Menten}, {Motte},
  {Sridharan}, \& {Wyrowski}}]{2002ApJ...566..945B}
{Beuther}, H., {Schilke}, P., {Menten}, K.~M., {et~al.} 2002, \apj, 566, 945

\bibitem[{{Beuther} {et~al.}(2005){Beuther}, {Sridharan}, \&
  {Saito}}]{2005ApJ...634L.185B}
{Beuther}, H., {Sridharan}, T.~K., \& {Saito}, M. 2005, \apjl, 634, L185

\bibitem[{{Bronfman} {et~al.}(1996){Bronfman}, {Nyman}, \&
  {May}}]{1996A&AS..115...81B}
{Bronfman}, L., {Nyman}, L.-A., \& {May}, J. 1996, \aaps, 115, 81

\bibitem[{{Carey} {et~al.}(1998){Carey}, {Clark}, {Egan}, {Price}, {Shipman},
  \& {Kuchar}}]{1998ApJ...508..721C}
{Carey}, S.~J., {Clark}, F.~O., {Egan}, M.~P., {et~al.} 1998, \apj, 508, 721

\bibitem[{{Carey} {et~al.}(2000){Carey}, {Feldman}, {Redman}, {Egan},
  {MacLeod}, \& {Price}}]{2000ApJ...543L.157C}
{Carey}, S.~J., {Feldman}, P.~A., {Redman}, R.~O., {et~al.} 2000, \apjl, 543,
  L157

\bibitem[{{Diolaiti} {et~al.}(2000){Diolaiti}, {Bendinelli}, {Bonaccini},
  {Close}, {Currie}, \& {Parmeggiani}}]{2000SPIE.4007..879D}
{Diolaiti}, E., {Bendinelli}, O., {Bonaccini}, D., {et~al.} 2000, in Presented
  at the Society of Photo-Optical Instrumentation Engineers (SPIE) Conference,
  Vol. 4007, Proc. SPIE Vol. 4007, p. 879-888, Adaptive Optical Systems
  Technology, Peter L. Wizinowich; Ed., ed. P.~L. {Wizinowich}, 879--888

\bibitem[{{Dobashi} {et~al.}(2005){Dobashi}, {Uehara}, {Kandori}, {Sakurai},
  {Kaiden}, {Umemoto}, \& {Sato}}]{2005PASJ...57S...1D}
{Dobashi}, K., {Uehara}, H., {Kandori}, R., {et~al.} 2005, \pasj, 57, 1

\bibitem[{{Draine} \& {Lee}(1984)}]{1984ApJ...285...89D}
{Draine}, B.~T. \& {Lee}, H.~M. 1984, \apj, 285, 89

\bibitem[{{Egan} {et~al.}(1998){Egan}, {Shipman}, {Price}, {Carey}, {Clark}, \&
  {Cohen}}]{1998ApJ...494L.199E}
{Egan}, M.~P., {Shipman}, R.~F., {Price}, S.~D., {et~al.} 1998, \apjl, 494,
  L199+

\bibitem[{{Froebrich} {et~al.}(2005){Froebrich}, {Ray}, {Murphy}, \&
  {Scholz}}]{2005A&A...432L..67F}
{Froebrich}, D., {Ray}, T.~P., {Murphy}, G.~C., \& {Scholz}, A. 2005, \aap,
  432, L67

\bibitem[{{Henning} {et~al.}(1990){Henning}, {Pfau}, \&
  {Altenhoff}}]{1990A&A...227..542H}
{Henning}, T., {Pfau}, W., \& {Altenhoff}, W.~J. 1990, \aap, 227, 542

\bibitem[{{Indebetouw} {et~al.}(2005){Indebetouw}, {Mathis}, {Babler}, {Meade},
  {Watson}, {Whitney}, {Wolff}, {Wolfire}, {Cohen}, {Bania}, {Benjamin},
  {Clemens}, {Dickey}, {Jackson}, {Kobulnicky}, {Marston}, {Mercer},
  {Stauffer}, {Stolovy}, \& {Churchwell}}]{2005ApJ...619..931I}
{Indebetouw}, R., {Mathis}, J.~S., {Babler}, B.~L., {et~al.} 2005, \apj, 619,
  931

\bibitem[{{Johnstone} {et~al.}(2003){Johnstone}, {Fiege}, {Redman}, {Feldman},
  \& {Carey}}]{2003ApJ...588L..37J}
{Johnstone}, D., {Fiege}, J.~D., {Redman}, R.~O., {Feldman}, P.~A., \& {Carey},
  S.~J. 2003, \apjl, 588, L37

\bibitem[{{Krumholz} \& {McKee}(2008)}]{2008Natur.451.1082K}
{Krumholz}, M.~R. \& {McKee}, C.~F. 2008, \nat, 451, 1082

\bibitem[{{Levine} {et~al.}(2008){Levine}, {Heiles}, \&
  {Blitz}}]{2008ApJ...679.1288L}
{Levine}, E.~S., {Heiles}, C., \& {Blitz}, L. 2008, \apj, 679, 1288

\bibitem[{{Lombardi}(2008)}]{2008arXiv0809.3383L}
{Lombardi}, M. 2008, ArXiv e-prints

\bibitem[{{Lombardi} \& {Alves}(2001)}]{2001A&A...377.1023L}
{Lombardi}, M. \& {Alves}, J. 2001, \aap, 377, 1023

\bibitem[{{Mathis}(1990)}]{1990ARA&A..28...37M}
{Mathis}, J.~S. 1990, \araa, 28, 37

\bibitem[{{Nyman} {et~al.}(2001){Nyman}, {Lerner}, {Nielbock}, {Anciaux},
  {Brooks}, {Chini}, {Albrecht}, {Lemke}, {Kreysa}, {Zylka}, {Johansson},
  {Bronfman}, {Kontinen}, {Linz}, \& {Stecklum}}]{2001Msngr.106...40N}
{Nyman}, L.-{\AA}., {Lerner}, M., {Nielbock}, M., {et~al.} 2001, The Messenger,
  106, 40

\bibitem[{{Ossenkopf} \& {Henning}(1994)}]{1994A&A...291..943O}
{Ossenkopf}, V. \& {Henning}, T. 1994, \aap, 291, 943

\bibitem[{{Perault} {et~al.}(1996){Perault}, {Omont}, {Simon}, {Seguin},
  {Ojha}, {Blommaert}, {Felli}, {Gilmore}, {Guglielmo}, {Habing}, {Price},
  {Robin}, {de Batz}, {Cesarsky}, {Elbaz}, {Epchtein}, {Fouque}, {Guest},
  {Levine}, {Pollock}, {Prusti}, {Siebenmorgen}, {Testi}, \&
  {Tiphene}}]{1996A&A...315L.165P}
{Perault}, M., {Omont}, A., {Simon}, G., {et~al.} 1996, \aap, 315, L165

\bibitem[{{Peretto} {et~al.}(2008){Peretto}, {Fuller}, {Andr{\'e}}, \&
  {Hennebelle}}]{2008ASPC..387...50P}
{Peretto}, N., {Fuller}, G.~A., {Andr{\'e}}, P., \& {Hennebelle}, P. 2008, in
  Astronomical Society of the Pacific Conference Series, Vol. 387, Astronomical
  Society of the Pacific Conference Series, ed. H.~{Beuther}, H.~{Linz}, \&
  T.~{Henning}, 50--+

\bibitem[{{Pillai} {et~al.}(2006{\natexlab{a}}){Pillai}, {Wyrowski}, {Carey},
  \& {Menten}}]{2006A&A...450..569P}
{Pillai}, T., {Wyrowski}, F., {Carey}, S.~J., \& {Menten}, K.~M.
  2006{\natexlab{a}}, \aap, 450, 569

\bibitem[{{Pillai} {et~al.}(2006{\natexlab{b}}){Pillai}, {Wyrowski}, {Menten},
  \& {Kr{\"u}gel}}]{2006A&A...447..929P}
{Pillai}, T., {Wyrowski}, F., {Menten}, K.~M., \& {Kr{\"u}gel}, E.
  2006{\natexlab{b}}, \aap, 447, 929

\bibitem[{{Rathborne} {et~al.}(2006){Rathborne}, {Jackson}, \&
  {Simon}}]{2006ApJ...641..389R}
{Rathborne}, J.~M., {Jackson}, J.~M., \& {Simon}, R. 2006, \apj, 641, 389

\bibitem[{{Rathborne} {et~al.}(2008){Rathborne}, {Jackson}, {Zhang}, \&
  {Simon}}]{2008arXiv0808.2973R}
{Rathborne}, J.~M., {Jackson}, J.~M., {Zhang}, Q., \& {Simon}, R. 2008, ArXiv
  e-prints, 808

\bibitem[{{Saito} {et~al.}(2001){Saito}, {Mizuno}, {Moriguchi}, {Matsunaga},
  {Onishi}, {Mizuno}, \& {Fukui}}]{2001PASJ...53.1037S}
{Saito}, H., {Mizuno}, N., {Moriguchi}, Y., {et~al.} 2001, \pasj, 53, 1037

\bibitem[{{Siegel} \& {Castellan}(1988)}]{1988psa..book.....S}
{Siegel}, S. \& {Castellan}, N.~J. 1988, {Nonparametric Statistics for the
  Behavioural Sciences} (McGraw-Hill)

\bibitem[{{Simon} {et~al.}(2006{\natexlab{a}}){Simon}, {Jackson}, {Rathborne},
  \& {Chambers}}]{2006ApJ...639..227S}
{Simon}, R., {Jackson}, J.~M., {Rathborne}, J.~M., \& {Chambers}, E.~T.
  2006{\natexlab{a}}, \apj, 639, 227

\bibitem[{{Simon} {et~al.}(2006{\natexlab{b}}){Simon}, {Rathborne}, {Shah},
  {Jackson}, \& {Chambers}}]{2006ApJ...653.1325S}
{Simon}, R., {Rathborne}, J.~M., {Shah}, R.~Y., {Jackson}, J.~M., \&
  {Chambers}, E.~T. 2006{\natexlab{b}}, \apj, 653, 1325

\bibitem[{{Sridharan} {et~al.}(2005){Sridharan}, {Beuther}, {Saito},
  {Wyrowski}, \& {Schilke}}]{2005ApJ...634L..57S}
{Sridharan}, T.~K., {Beuther}, H., {Saito}, M., {Wyrowski}, F., \& {Schilke},
  P. 2005, \apjl, 634, L57

\bibitem[{{Steinacker} {et~al.}(2005){Steinacker}, {Bacmann}, {Henning},
  {Klessen}, \& {Stickel}}]{2005A&A...434..167S}
{Steinacker}, J., {Bacmann}, A., {Henning}, T., {Klessen}, R., \& {Stickel}, M.
  2005, \aap, 434, 167

\bibitem[{{Wall} \& {Jenkins}(2003)}]{2003psa..book.....W}
{Wall}, J.~V. \& {Jenkins}, C.~R. 2003, {Practical Statistics for Astronomers}
  (Princeton Series in Astrophysics)

\bibitem[{{Weferling} {et~al.}(2002){Weferling}, {Reichertz}, {Schmid-Burgk},
  \& {Kreysa}}]{2002A&A...383.1088W}
{Weferling}, B., {Reichertz}, L.~A., {Schmid-Burgk}, J., \& {Kreysa}, E. 2002,
  \aap, 383, 1088

\bibitem[{{Weingartner} \& {Draine}(2001)}]{2001ApJ...548..296W}
{Weingartner}, J.~C. \& {Draine}, B.~T. 2001, \apj, 548, 296

\bibitem[{{Zinnecker} \& {Yorke}(2007)}]{2007ARA&A..45..481Z}
{Zinnecker}, H. \& {Yorke}, H.~W. 2007, \araa, 45, 481

\end{thebibliography}

\end{document}